\documentclass[aps,prd,superscriptaddress,nofootinbib,10pt,colorlinks,linkcolor=red,anchorcolor=green,citecolor=blue]{revtex4-2}
\usepackage{float} 
\usepackage{diagbox}
\usepackage{graphicx}
\usepackage{bm}
\usepackage{amsmath}
\usepackage{mathrsfs}
\usepackage{ulem}
\usepackage{color}
\usepackage{amsthm,amsmath,amssymb}
\usepackage{inputenc}
\usepackage{verbatim}
\usepackage{siunitx} 
\usepackage{booktabs}
\usepackage{hyperref}
\usepackage{orcidlink}
\usepackage{xcolor}
\begin{document}

\title{Analytical solution and Lie algebra of the relativistic Boltzmann equation}
\author{Yi Wang\,\orcidlink{0009-0004-4811-2391}}
\affiliation{Department of Physics, Tsinghua University, Beijing 100084, China}
\author{Xuan Zhao\,\orcidlink{0009-0007-6093-5464}}
\affiliation{Department of Physics, Tsinghua University, Beijing 100084, China}
\author{Zhe Xu}
\affiliation{Department of Physics, Tsinghua University, Beijing 100084, China}
\author{Jin Hu\,\orcidlink{0000-0003-1179-4603}}
\email[]{hu-j23@fzu.edu.cn}
\affiliation{Department of Physics, Fuzhou University, Fujian 350116, China}


\begin{abstract}
In this work, we present a novel and more efficient approach to constructing the relativistic Bobylev, Krook, and Wu (BKW) solution [A. V. Bobylev, Sov. Phys. Dokl. \textbf{20}, 820 (1976); M. Krook and T. T. Wu,  Phys. Fluids \textbf{20}, 1589 (1977)] of the relativistic Boltzmann equation. Introducing a new ansatz for the distribution function, we demonstrate that within this specific ansatz space, only the equilibrium and BKW-type forms yield exact solutions to the nonlinear Boltzmann equation. Furthermore, we also derive the Lie algebra of invariant transformations admitted by the relativistic Boltzmann equation and show that the corresponding symmetry group transformations can be systematically constructed from this algebra.
\end{abstract}

\maketitle

\section{Introduction} 
\label{Introduction}
The relativistic Boltzmann equation serves as a fundamental framework for describing the kinetic behavior of dilute relativistic gases in nonequilibrium regimes and occupies a central role in relativistic kinetic theory. It provides a statistical description of how the distribution function of particles evolves under the influence of free streaming and particle collisions, making it particularly suitable for systems where deviations from local thermal equilibrium are significant. Due to its solid theoretical foundation, the equation has found broad applications across diverse domains of modern physics, ranging from high-energy heavy-ion collisions \cite{Heinz:1984yq,Heinz:1985qe,Bass:1998ca,Arnold:2000dr,Molnar:2001ux,Xu:2004mz,Denicol:2012cn} to astrophysical phenomena and cosmology \cite{Dodelson:2003ft,Weinberg:2008zzc}. In these contexts, it enables the study of nonequilibrium transport processes under extreme conditions, such as the quark-gluon plasma produced in ultrarelativistic heavy-ion collisions and in the early universe, thereby offering a crucial link between microscopic dynamics and macroscopic observables.

In 1872, Boltzmann first formulated a kinetic equation describing the time evolution of the single-particle distribution function $f$ for dilute gas molecules \cite{boltzmann1872boltzmann}, establishing the H-theorem to demonstrate the monotonic decrease of the H function in isolated systems, which provided a microscopic foundation for the macroscopic law of entropy production. Due to the foundational role of the Boltzmann equation in statistical physics, the search for its analytical solutions has remained a central task for theoretical physicists since then. However, the mathematical complexity of the collision integral and the strong nonlinearity inherent in the equation make the derivation of exact analytical solutions extremely difficult. For nearly a century thereafter, the only well-known exact solution of the Boltzmann equation was the equilibrium-state solution. From the perspective of solving the equation, this solution is undoubtedly trivial, as it can be directly constructed from the moment the equation is written, based on the principle of collisional invariance. This situation remained unchanged until the discovery of the Bobylev, Krook, and Wu (BKW) solution, the first nontrivial exact solution to the Boltzmann equation.  

In 1976, Bobylev discovered a set of particular solutions from the self-similar solutions of the Fourier-transformed Boltzmann equation \cite{Bobylev1}. Almost simultaneously, Krook and Wu also found this particular solution using different methods \cite{KW2,KW}, leading to its designation as the BKW solution. \footnote{This solution was originally derived by R. S. Krupp in his master's thesis in 1967 \cite{dissertation}, predating its later rediscovery as reviewed by \cite{exactreview}.} This solution describes the nonlinear relaxation dynamics of a spatially homogeneous, nonrelativistic gas, offering a nice interpretation  of the behavior of the one-particle distribution function at the high-momentum tail of the spectrum \cite{KW2}. The BKW solution is not only a paradigmatic example of nonequilibrium relaxation but also provides deep insight into the interplay between collisional dissipation and the restoration of thermodynamic equilibrium. Due to its exact analytical form, it serves as an indispensable benchmark for rigorously testing theoretical approximations\textemdash such as moment methods and model kinetic equations\textemdash as well as numerical schemes like the direct simulation Monte Carlo method \cite{bird1994molecular} and other numerical Boltzmann solvers \cite{Xu:2004mz}, particularly in far-from-equilibrium regimes. Furthermore, the BKW solution has been further extended to relativistic regimes recently, including expanding solutions in Friedmann-Lema{\^i}tre-Robertson-Walker spacetime \cite{Bazow:2015dha,Bazow:2016oky} and solutions with anisotropic scattering cross sections \cite{Hu:2024utr}. In this work, we employ a novel approach to reconstruct the relativistic BKW solution.

Given the known exact solutions to the Boltzmann equation, a natural question emerges: whether it is possible to construct or discover additional exact solutions of physical significance by building upon these known solutions. Historically, this question is intimately tied to the Lie group analysis of the partial differential equation (PDE). From a mathematical perspective, a central theme is the classification of PDE solutions, where it is often found that solutions with apparently different forms are connected via nontrivial symmetry transformations. This idea is intuitively compelling and can be rigorously formulated using Lie group and Lie algebra methods: any transformation that leaves the form of the original PDE invariant will necessarily map one solution to another. Let us now return to a concrete discussion of the Boltzmann equation. As a complex integro-differential equation, it can still be subjected to Lie group and Lie algebra analysis. Such a symmetry-based approach offers a promising pathway to discovering new physically relevant exact solutions. Even if new desired solutions are not immediately found, this framework may help establish connections between existing ones, thereby uncovering deeper physical relationships hidden within the structure of the equation hopefully. Although the Lie algebra of the Boltzmann equation is of significant importance and could offer valuable insights, its systematic and comprehensive exploration  within the relativistic kinetic theory has been lacking. Within this work, we aim to close this gap.

The paper is structured as follows: In Sec.~\ref{Lie Groups, Lie Algebras, and Exact Analytical Solutions for the Boltzmann Equation}, we provide a brief introduction to the relativistic Boltzmann equation, and reconstruct the BKW solution using an alternative yet straightforward approach. Section~\ref{Extended Symmetry Generators for the Relativistic Boltzmann Equation and Applications} is devoted to determining the Lie algebras admitted by the equation. The summary and outlook are given in Sec.~\ref{Conclusion and outlook}. Natural units $k_{\mathrm{B}} = c = \hbar = 1$ are used. The metric tensor is represented by $g_{\mu \nu} = \mathrm{diag} \left( 1,-1,-1,-1 \right)$. We also use the abbreviation $\mathrm{d}P$ to denote the Lorentz invariant integral measure for on-shell massless particles $\int \mathrm{d}P \equiv \frac{2}{\left(2 \pi \right)^3} \int \mathrm{d}^4 p \theta \left( p^0 \right) \delta \left( p^2 \right)$. For such particles, the massless condition $p^\mu p_\mu = m_0^2 \rightarrow 0$ implies  $p^0 = p = |\boldsymbol{p}|$.

\section{Boltzmann equation and reconstruction of the BKW solution} 
\label{Lie Groups, Lie Algebras, and Exact Analytical Solutions for the Boltzmann Equation}

\subsection{Boltzmann equation}
\label{Boltzmann equation}
In the absence of external fields, the spacetime evolution of a nonequilibrium Bose or Fermi gas is described by the relativistic Boltzmann equation
\begin{align} \begin{split} p \cdot \partial f \left( x^{\mu} ,p^{\mu} \right) = C\left[f\right], \label{21} \end{split} \end{align}
with the collision kernel defined as
\begin{align} \begin{split} C \left[ f \right]  \equiv &\frac{1}{2} \int \mathrm{d}K \mathrm{d}P_f \mathrm{d}K_f \left[ f \left( x^{\mu}, p_f^{\mu} \right) f \left( x^{\mu}, k_f^{\mu} \right) \left( 1 + a f \left( x^{\mu}, p^{\mu} \right) \right) \left( 1 + a f \left( x^{\mu}, k^{\mu} \right) \right) \right.\\
                 &\left. - f \left( x^{\mu}, p^{\mu} \right) f \left( x^{\mu}, k^{\mu} \right) \left( 1 + a f \left( x^{\mu}, p_f^{\mu} \right) \right) \left( 1 + a f \left(x^{\mu}, k_f^{\mu} \right) \right) \right] W_{p,k \to p_f,k_f}, \label{22} \end{split} \end{align}
\noindent where $f \left( x^{\mu}, p^{\mu} \right)$ is the single-particle distribution function in phase space and only binary elastic collisions are taken into account. The parameter $a$ is a coefficient labeling different statistics with $a = 1 \left( -1 \right)$ corresponding to the effect of Bose enhancement (Fermi blocking), while $a = 0$  follows from the classical Maxwell-Boltzmann statistics. The transition rate $W_{p,k \to p_f,k_f}$ can be expressed as \cite{DeGroot:1980dk}
\begin{equation}W_{p,k \to p_f,k_f} = \left(2 \pi \right)^6 s \sigma \left( s, \Theta \right) \delta^4 \left( p^\mu + k^\mu - p_f^\mu - k_f^\mu \right)\end{equation} 
with the differential cross section $\sigma \left( s, \Theta \right)$ depending on the Mandelstam variable $s \equiv \left( p + k \right)^2$ and the scattering angle $\Theta$,
\begin{equation}
    \cos \Theta \equiv \frac{(p-k) \cdot (p_f-k_f)}{(p-k)^2}.
\end{equation} 
For convenience, in the following we assume that the angular dependence of the differential cross section is solely through the cosine of the scattering angle, i.e.,  $\cos \Theta \equiv \kappa$.  Accordingly, the differential cross section is expressed as $\sigma \left( s, \kappa \right)$ with . Note that Eq.\eqref{21} applies specifically to single-component systems; for extensions to multicomponent systems, see \cite{DeGroot:1980dk, Hu:2022vph}. For the majority of this work, we focus on the case $a = 0$ , i.e., the Boltzmann equation under classical Maxwell Boltzmann statistics, including a rederivation of the BKW solution using a novel  method. In the following sections, we will see that, when analyzing the symmetries of the equation, the classical Boltzmann equation exhibits a richer symmetry structure compared to its quantum counterparts. Informally speaking, it is ``more symmetric.''

When reconstructing exact analytical solutions of the Boltzmann equation, we proceed by imposing the assumptions of spatial homogeneity, i.e., $f = f \left( x^0, p^{\mu} \right)$. Subsequently, Eq.(\ref{21}) becomes
\begin{align} \begin{split} p^0 \partial_0 f \left( x^0, p^{\mu} \right) = C \left[ f \right], \label{23} \end{split} \end{align}
\noindent where $C \left[ f \right]$ can be rewritten as
\begin{align} \begin{split} C \left[ f \right] & = \frac{ \left( 2 \pi \right)^6}{2} \int \mathrm{d}K \mathrm{d}P_f \mathrm{d}K_f s \sigma \left( s, \kappa \right) \delta^4 \left( p^\mu + k^\mu - p_f^\mu - k_f^\mu \right) 
                \left( f \left( x^0, p_f^{\mu} \right) f \left( x^0, k_f^{\mu} \right) - f \left( x^0, p^{\mu} \right) f \left( x^0, k^{\mu} \right) \right), \label{24} \end{split} \end{align}
where the particles are taken to be massless so that $p^0 = p = |\boldsymbol{p}|$. Unless otherwise specified, we assume throughout that the particle mass is zero. 

For later convenience, we present here a frequently used parametrization of the cross section in our following discussion
\begin{align} \begin{split} \sigma \left( s, \kappa \right) = s^{l - 1} g_l \left( \kappa \right). \label{25} \end{split} \end{align}    
We will see that this separable form of the cross section\textemdash depending on the Mandelstam variable $s$ and the scattering angle independently\textemdash plays a crucial role in analyzing the scaling behavior of the Boltzmann equation. In fact, our consideration extends beyond this specific form: any cross section that respects a scale transformation of the form $\sigma \rightarrow \lambda^w \sigma$ [in the case of Eq.\eqref{25}, $w = 2 \left( l - 1 \right)$] under momentum rescaling $p^\mu \rightarrow \lambda p^\mu$ falls within the scope of our interest. A more comprehensive discussion of such scale-covariant cross sections will be provided in the following sections. While this unified mathematical form explicitly separates the angular dependence from the energy dependence, it physically encompasses standard collision models extensively discussed in classic kinetic theory texts (see, e.g., \cite{DeGroot:1980dk, Denicol:2021icf}). Specifically, it is rather straightforward to see that this parametrization includes two widely encountered scattering interactions as special cases: when $g_l \left( \kappa \right) = {\rm const}$, $l = 1$ corresponds to the hard-sphere interaction, while $l = 0$ corresponds to the leading order scalar $\phi^4$ theory.

\subsection{Known closed-form solutions}
\label{known}
In general, directly solving for the generators and their corresponding transformations from the definition is a cumbersome mathematical task. The standard Lie symmetry analysis typically leads to a set of highly involved, nonlinear PDEs, which are often prohibitively difficult to solve explicitly (see Sec.~\ref{Extended Symmetry Generators for the Relativistic Boltzmann Equation and Applications}). Rather than pursuing brute-force computation, we adopt a physics-informed strategy, guided by accumulated knowledge or physical insights in kinetic theory, to determine as much as possible about the relevant Lie algebra. For instance, knowledge of the various closed-form solutions of the Boltzmann equation offers helpful insights that guide the construction of the Lie algebra. Below, we briefly review and collect the known exact solutions for reference and further analysis.

When the Boltzmann equation is formulated, it is already known that the equilibrium solution constitutes a special class of exact solutions. These solutions are of fundamental physical significance, as they form the basis of equilibrium statistical mechanics. Based on the principle of collisional invariance in microscopic scattering processes, they can be systematically constructed \cite{DeGroot:1980dk}. Depending on the value of the parameter $a$, these equilibrium distributions take the following unified form:
\begin{align} f \left(p^\mu, u^\mu, \mu_c, T \right) = \frac{1}{e^{\frac{E_p - \mu_c}{T}} - a}, \label{three equilibrium distributions} \end{align}
\noindent where $\mu_c$ is the chemical potential, $T$ is the temperature, and $E_p \equiv u \cdot p$ with a timelike four-vector $u^\mu$ which is often interpreted as the fluid velocity applied to a fluid system. Here, $a=1$, $-1$, and $0$ correspond to Bose-Einstein, Fermi-Dirac, and classical Maxwell-Boltzmann statistics, respectively. In the above expressions, we have suppressed the normalization factors for simplicity.

The second solution is the nonthermal fixed point (NTFP) solution to the spatially homogeneous and momentum-isotropic Boltzmann equation, which has attracted significant research interest in recent years \cite{Berges:2008wm,Berges:2008mr}. In the over-occupied regime of a Bose gas, where the distribution function $f \gg 1$, the statistical factors in the collision term of the Boltzmann equation become
\begin{align}
    &f \left( t, p_f^0 \right) f \left( t, k_f^0 \right) \left( 1 + f \left( t, p^0 \right) \right) \left( 1 +  f \left( t, k^0 \right) \right) - f \left( t, p^0 \right) f \left( t, k^0 \right) \left( 1 + f \left( t, p_f^0 \right) \right) \left( 1 + f \left( t, k_f^0 \right) \right) \nonumber \\
    \approx\,\, & f \left( t, p_f^0 \right) f \left( t, k_f^0 \right) \left( f \left( t, p^0 \right) + f \left( t, k^0 \right) \right) - f \left( t, p^0 \right) f \left( t, k^0 \right) \left( f \left( t, p_f^0 \right) + f \left( t, k_f^0 \right) \right). \label{29}
    \end{align}
The above simplification is crucial to the analysis of the scaling behavior of NTFP, and we shall come to this point later. The Boltzmann equation with the replacement of Eq.\eqref{29} admits a universal self-similar solution; i.e., the transport in isolated systems is well described in terms of a self-similar evolution,
\begin{align} \begin{split}
                f \left( t, p^0 \right) & = t^{\tilde{a}} f_s \left( \xi \equiv t^{\tilde{b}} p^0 \right), \label{210} \end{split} \end{align}
in a given scaling regime. The scaling exponents $\tilde{a}$ and $\tilde{b}$, as well as the functional form of the nonthermal fixed point distribution $f_s \left( \xi \right)$, are universal \cite{PineiroOrioli:2015cpb}. The above solution exhibits a typical inverse particle cascade, which leads to Bose condensation in the highly occupied low-momentum regime. 
This particle transport toward low momenta is part of a dual cascade, in which energy is simultaneously transferred to higher momenta via a direct cascade. In this regime, distinct scaling exponents ${\tilde{a}}^\prime,{\tilde{b}}^\prime$ and a different scaling function $f^\prime_s$ as compared to the infrared regime are observed. We refer interested readers to \cite{Berges:2008wm,Berges:2008mr,PineiroOrioli:2015cpb} for further technical details. Below, we would like to give a few remarks:
\begin{itemize}
  \item First, strictly speaking, the NTFP is not an analytical solution, as the explicit form of $f_s$ cannot be determined analytically. Nor can it be considered an exact solution of the full Boltzmann equation, as it relies on simplifications or approximations such as Eq.\eqref{29}. Nevertheless, it captures an approximately universal dynamical behavior, making it physically compelling. Notably, this self-similar solution is relevant for a wide range of applications from ultracold quantum gases to high-energy particle physics, and the existence of nonthermal fixed points acts as a nonequilibrium attractor for isolated many-body systems far from equilibrium \cite{Mikheev:2023juq}.
  \item Second, the NTFP is a fixed point distinct from thermal equilibrium. Its existence implies that the fate of a weakly coupled system’s evolution need not be thermalization\textemdash within physically relevant timescales. For instance, in the infrared regime, the inverse particle cascade eventually leads to the onset of Bose condensation.
  \item Third, the NTFP may share a deep connection with another exact analytical solution of the  Boltzmann equation with classical statistics to be discussed later. At first glance, both are self-similar solutions; however, a more profound relationship may exist, which goes beyond superficial similarity and warrants further investigation. However, it should be noted that only the behavior of the NTFP solution in the ultraviolet regime is truly comparable, as in this regime the scaling requires $f \ll 1$, causing the transport equation to reduce to the classical Boltzmann equation.
\end{itemize}

The last exact analytical solution is the relativistic BKW solution of the Boltzmann equation. Notably, this nonequilibrium particular solution possesses a fixed point that corresponds to the equilibrium solution of the Boltzmann equation \cite{Hu:2024utr}.

\subsection{Reconstruction of the BKW solution}
\label{Reconstruction of the BKW solution}
In this subsection, we will construct the relativistic BKW solution using a novel approach that is simpler and more efficient than the conventional moment method~\cite{Bazow:2015dha,Bazow:2016oky,Hu:2024utr}, as it only requires evaluating the collision integral and solving a system of ordinary differential equations. It should be noted that the relativistic BKW solution applies only to the case where $\sigma \left( s, \kappa \right) = g_l \left( \kappa \right)$ \cite{Hu:2024utr}, corresponding to $l=1$ in Eq.\eqref{25}. For simplicity, we consider the hard-sphere interaction, i.e.,  $\sigma \left( s, \kappa \right) \equiv \sigma = \text{const}$.
Before proceeding to the explicit calculation, let us first simplify the collision integral as much as possible \cite{Arnold:2003,Soudi:2021}.

It is important to emphasize that when deriving the BKW solution, we work exclusively in the local rest frame of the homogeneous and isotropic gas. In this frame, the macroscopic flow velocity is defined as $u^\mu = (1, 0, 0, 0)$. Consequently, for massless particles, the energy simplifies to $p^0 = p \cdot u = |\boldsymbol{p}|$, and the distribution function $f(x^0, p^0)$ depends only on $x^0$ and $p^0$. All subsequent phase-space integrals are evaluated in this chosen frame. The collision term in Eq.\eqref{24} can be written as
\begin{align} \begin{split} C \left[ f \right] & = \frac{\sigma_\mathrm{T}}{2 \left( 2 \pi \right)^4}  \int \frac{\mathrm{d}^3 \boldsymbol{k}}{k^0} \frac{\mathrm{d}^3 \boldsymbol{p}_f}{p_f^0} \frac{\mathrm{d}^3 \boldsymbol{k}_f}{k_f^0} s \delta^4 \left( p^\mu + k^\mu - p_f^\mu - k_f^\mu \right) \left( f \left( x^0, p_f^0 \right) f \left( x^0, k_f^0 \right) - f \left( x^0, p^0 \right) f \left( x^0, k^0 \right) \right), \label{211} \end{split} \end{align}
where $\sigma_\mathrm{T} \equiv 2 \pi \sigma$ is the total cross section. We can reexpress the phase-space integral by introducing the definitions of the three-momentum difference $\boldsymbol{q} \equiv \boldsymbol{p} - \boldsymbol{p}_f = \boldsymbol{k}_f - \boldsymbol{k}$ and the energy difference $\omega \equiv p^0 - p_f^0 = k_f^0 - k^0$. Then the integral over the delta function can be cast into
\begin{align} \begin{split} & \int \frac{\mathrm{d}^3 \boldsymbol{k}}{k^0} \frac{\mathrm{d}^3 \boldsymbol{p}_f}{p_f^0} \frac{\mathrm{d}^3 \boldsymbol{k}_f}{k_f^0} \delta^4 \left( p^\mu + k^\mu - p_f^\mu - k_f^\mu \right) \\
                & = \int \frac{\mathrm{d}^3 \boldsymbol{k}}{k^0} \frac{\mathrm{d}^3 \boldsymbol{p}_f}{p_f^0} \frac{\mathrm{d}^3 \boldsymbol{k}_f}{k_f^0} \delta^3 \left( \boldsymbol{p} + \boldsymbol{k} - \boldsymbol{p}_f - \boldsymbol{k}_f \right) \int \mathrm{d} \omega \delta \left( p^0 - \omega - p_f^0 \right) \delta \left( k^0 + \omega - k_f^0 \right) \\
                & = \int \frac{\mathrm{d}^3 \boldsymbol{k}}{k^0} \frac{\mathrm{d}^3 \boldsymbol{p}_f}{p_f^0} \frac{\mathrm{d}^3 \boldsymbol{k}_f}{k_f^0} \mathrm{d} \omega \delta^3 \left( \boldsymbol{p} + \boldsymbol{k} - \boldsymbol{p}_f - \boldsymbol{k}_f \right)  \delta \left( p^0 - \omega - p_f^0 \right) \delta \left( k^0 + \omega - k_f^0 \right) \\
                & = \int \mathrm{d}^3 \boldsymbol{k} \mathrm{d}^3 \boldsymbol{q} \mathrm{d} \omega \frac{1}{|\boldsymbol{k}| |\boldsymbol{p}_f| |\boldsymbol{q+k}|}  \delta \left( |\boldsymbol{p}| - \omega - |\boldsymbol{p} - \boldsymbol{q}| \right) \delta \left( |\boldsymbol{k}| + \omega - |\boldsymbol{k} + \boldsymbol{q}| \right)\\
                & = \int \mathrm{d}^3 \boldsymbol{k} \mathrm{d}^3 \boldsymbol{q} \mathrm{d} \omega \frac{1}{|\boldsymbol{k}| |\boldsymbol{p}_f| |\boldsymbol{q+k}|}  \delta \left( |\boldsymbol{p}_f| - \left( |\boldsymbol{p}| - \omega \right) \right) \delta \left( |\boldsymbol{k}_f| - \left( \omega + |\boldsymbol{k}| \right) \right).\label{212} \end{split} \end{align}
Here, we continue to use the on-shell relation for massless particles: $p^0 = p = |\boldsymbol{p}|$. By further simplifying the collision integral, with the detailed evaluation  relegated to Appendix \ref{AppendixCollision}, we arrive at the following explicit form:
\begin{align} \begin{split} C \left[ f \right]  = &\frac{\sigma_\mathrm{T}}{2 \left( 2 \pi \right)^4}  \int \mathrm{d}^3 \boldsymbol{k} \mathrm{d}^3 \boldsymbol{q} \mathrm{d} \omega \frac{1}{|\boldsymbol{p}| |\boldsymbol{k}|^2 |\boldsymbol{q}|^2} s    \delta \left( \cos \theta_{\boldsymbol{pq}} - \frac{\omega}{|\boldsymbol{q}|} + \frac{\omega^2 - |\boldsymbol{q}|^2}{2|\boldsymbol{p}||\boldsymbol{q}|} \right) \delta \left( \cos \theta_{\boldsymbol{kq}} - \frac{\omega}{|\boldsymbol{q}|} - \frac{\omega^2 - |\boldsymbol{q}|^2}{2|\boldsymbol{k}||\boldsymbol{q}|} \right) \\
                & \times \theta \left( |\boldsymbol{p}| - \omega \right) \theta \left( \omega + |\boldsymbol{k}| \right) \theta \left( |\boldsymbol{q}| - |\omega| \right) \theta \left( |\boldsymbol{p}| - \frac{|\boldsymbol{q}| + \omega}{2} \right) \theta \left( |\boldsymbol{k}| - \frac{|\boldsymbol{q}| - \omega}{2} \right) \\
                & \times \left( f \left( x^0, p_f^0 \right) f \left( x^0, k_f^0 \right) - f \left( x^0, p^0 \right) f \left( x^0, k^0 \right) \right), \label{219} \end{split} \end{align}
where $\theta_{\boldsymbol{pq}}$ and $\theta_{\boldsymbol{kq}}$ denote the angles between $\boldsymbol{q}$ and the momenta $\boldsymbol{p}$ and $\boldsymbol{k}$, respectively.

The preparatory steps are now complete. In the following, we will demonstrate how to construct the BKW solution using the new method. We begin by introducing two dimensionless notations $\tau \equiv T^3 \sigma_\mathrm{T} x^0$ and  $\hat{p}^\mu \equiv \frac{p^\mu}{T}$ where $T$  is a typical energy scale and can be identified with the temperature in certain physical scenarios, and we will omit the hat notation from $\hat{p}^\mu$ when nothing confusing occurs. Then the dimensionless Boltzmann equation takes the form of
\begin{align} p^0 \partial_{\tau} f \left( \tau, p^0 \right) = C \left[ f \right], \label{220} \end{align}
with
\begin{align} \begin{split} C \left[ f \right]  = & \frac{1}{2 \left( 2 \pi \right)^4} \int \mathrm{d}^3 \boldsymbol{k} \mathrm{d}^3 \boldsymbol{q} \mathrm{d} \omega \frac{1}{|\boldsymbol{p}| |\boldsymbol{k}|^2 |\boldsymbol{q}|^2} s   \delta \left( \cos \theta_{\boldsymbol{pq}} - \frac{\omega}{|\boldsymbol{q}|} + \frac{\omega^2 - |\boldsymbol{q}|^2}{2|\boldsymbol{p}||\boldsymbol{q}|} \right) \delta \left( \cos \theta_{\boldsymbol{kq}} - \frac{\omega}{|\boldsymbol{q}|} - \frac{\omega^2 - |\boldsymbol{q}|^2}{2|\boldsymbol{k}||\boldsymbol{q}|} \right) \\
                & \times \theta \left( |\boldsymbol{p}| - \omega \right) \theta \left( \omega + |\boldsymbol{k}| \right) \theta \left( |\boldsymbol{q}| - |\omega| \right) \theta \left( |\boldsymbol{p}| - \frac{|\boldsymbol{q}| + \omega}{2} \right) \theta \left( |\boldsymbol{k}| - \frac{|\boldsymbol{q}| - \omega}{2} \right) \left( f \left( \tau, p_f^0 \right) f \left( \tau, k_f^0 \right) - f \left( \tau, p^0 \right) f \left( \tau, k^0 \right) \right). \label{221} \end{split} \end{align}

We propose a class of ansatz distribution function with the following parametrization:
\begin{align} \begin{split} f \left( \tau, p^0 \right) = e^{- \frac{p^0}{\alpha \left( \tau \right)}} \sum_{i=0}^n A_i \left( \tau \right) \cdot \left( p^0 \right)^i, \label{222} \end{split} \end{align}
\noindent where $A_i \left( \tau \right)$ are unknown functions of time $\tau$ to be determined. Although this trial function does not encompass all possible parametrizations\textemdash hence it is not fully general\textemdash it is sufficiently flexible for our purposes. The proposed ansatz is not arbitrary, but guided by three physical and mathematical principles.
First, to ensure analytical tractability, the distribution function is assumed to be a summation over functions separable in its variables, i.e., their dependence on $\tau$ and $p^0$ is factorized. This restricts its functional form to a series expansion in functions of $p^0$, with coefficients that depend only on $\tau$. 
Second, to guarantee regularity, ensuring that all kinetic moments are finite, a natural choice is to introduce an extra regulating function that suppresses potential ultraviolet divergences. Third, we invoke the principle of minimal complexity: in the absence of additional symmetry or dynamical constraints, the most economical choice is to expand the $p^0$-dependent part as a power series in $p^0$, while taking the regulating function to be exponentially decaying (e.g., $e^{-\frac{p^0}{\alpha \left( \tau \right)}}$). Taken together, these considerations lead to the form adopted in this work. 

Substituting the above ansatz into both sides of the Boltzmann equation, we can determine the coefficients $A_i$ by the method of undetermined coefficients\textemdash provided that the powers of $p^0$  on the left- and right-hand sides match term by term. It is evident that not all values of $n$ can satisfy this condition. One can count the powers of $p^0$ on both sides of the equation: in the absence of subtle cancellations, the degrees no longer match for $n>1$ (we will discuss these cases in Appendix \ref{Trial function}).
Only two exceptions exist: $n=0$ corresponds to the equilibrium solution, while $n=1$ yields a solution of BKW type. This may partly explain why the BKW solution is so special. Focusing on the case of $n = 1$, Eq.\eqref{222} takes the following form:
\begin{align} \begin{split} f \left( \tau, p^0 \right) = e^{- \frac{p^0}{\alpha \left( \tau \right)}} \left( A_0 \left( \tau \right) + A_1 \left( \tau \right) p^0 \right). \label{223} \end{split} \end{align}

\noindent Substituting it into the left-hand side of Eq.\eqref{220} leads us to
\begin{align} \begin{split} e^{- \frac{p^0}{\alpha \left( \tau \right)}} \left( A_0' \left( \tau \right) p^0 + \left( A_1' \left( \tau \right) + \frac{A_0 \left( \tau \right) \alpha' \left( \tau \right)}{\alpha^2 \left( \tau \right)} \right) \left( p^0 \right)^2 + \frac{A_1 \left( \tau \right) \alpha' \left( \tau \right)}{\alpha^2 \left( \tau \right)} \left( p^0 \right)^3 \right), \label{224} \end{split} \end{align}
\noindent where  the  prime denotes the derivative of $A \left( \tau \right)$ with respect to $\tau$. Then the substitution of Eq.\eqref{223} into Eq.\eqref{221} gives
\begin{align} \begin{split} C \left[ f \right] = & \frac{1}{2 \left( 2 \pi \right)^4} \int \mathrm{d}^3 \boldsymbol{k} \mathrm{d}^3 \boldsymbol{q} \mathrm{d} \omega \frac{1}{|\boldsymbol{p}| |\boldsymbol{k}|^2 |\boldsymbol{q}|^2}e^{- \frac{|\boldsymbol{p}| + |\boldsymbol{k}|}{\alpha \left( \tau \right)}} A_1^2 \left( \tau \right) \delta \left( \cos \theta_{\boldsymbol{p}\boldsymbol{q}} - \frac{\omega}{|\boldsymbol{q}|} + \frac{\omega^2 - |\boldsymbol{q}|^2}{2|\boldsymbol{p}||\boldsymbol{q}|} \right) \delta \left( \cos \theta_{\boldsymbol{k}\boldsymbol{q}} - \frac{\omega}{|\boldsymbol{q}|} - \frac{\omega^2 - |\boldsymbol{q}|^2}{2|\boldsymbol{k}||\boldsymbol{q}|} \right) \\
                & \times \theta \left( |\boldsymbol{p}| - \omega \right) \theta \left( \omega + |\boldsymbol{k}| \right) \theta \left( |\boldsymbol{q}| - |\omega| \right) \theta \left( |\boldsymbol{p}| - \frac{|\boldsymbol{q}| + \omega}{2} \right) \theta \left( |\boldsymbol{k}| - \frac{|\boldsymbol{q}| - \omega}{2} \right) \\
                & \times \left( \omega \left( |\boldsymbol{p}| - |\boldsymbol{k}| \right) - \omega^2 \right) 2 |\boldsymbol{p}| |\boldsymbol{k}| \left( 1 - \left( \cos \theta_{\boldsymbol{p}\boldsymbol{q}} \cos \theta_{\boldsymbol{k}\boldsymbol{q}} + \cos \left( \phi_{\boldsymbol{p}} - \phi_{\boldsymbol{k}} \right) \sin \theta_{\boldsymbol{p}\boldsymbol{q}} \sin \theta_{\boldsymbol{kq}} \right) \right). \label{225} \end{split} \end{align}
\noindent In obtaining the above expression, we use $s = 2 |\boldsymbol{p}| |\boldsymbol{k}| \left( 1 - \cos \theta_{\boldsymbol{p}\boldsymbol{k}} \right)$ and $\cos \theta_{\boldsymbol{p}\boldsymbol{k}} = \cos \theta_{\boldsymbol{p}\boldsymbol{q}} \cos \theta_{\boldsymbol{k}\boldsymbol{q}} + \cos \left( \phi_{\boldsymbol{p}} - \phi_{\boldsymbol{k}} \right) \sin \theta_{\boldsymbol{p}\boldsymbol{q}} \sin \theta_{\boldsymbol{kq}}$.

By transforming the integral measure to spherical coordinates, Eq.\eqref{225} can be rearranged to yield
\begin{align} \begin{split} C \left[ f \right] = & \frac{1}{ \left( 2 \pi \right)^4} e^{- \frac{|\boldsymbol{p}|}{\alpha \left( \tau \right)}} A_1^2 \left( \tau \right) \int \mathrm{d}|\boldsymbol{k}| \mathrm{d} \cos \theta_{\boldsymbol{kq}} \mathrm{d} \phi_{\boldsymbol{k}} \mathrm{d}|\boldsymbol{q}| \mathrm{d} \cos \theta_{\boldsymbol{pq}} \mathrm{d} \phi_{\boldsymbol{q}} \mathrm{d} \omega \\
                & \times |\boldsymbol{k}| \delta \left( \cos \theta_{\boldsymbol{pq}} - \frac{\omega}{|\boldsymbol{q}|} + \frac{\omega^2 - |\boldsymbol{q}|^2}{2|\boldsymbol{p}||\boldsymbol{q}|} \right) \delta \left( \cos \theta_{\boldsymbol{kq}} - \frac{\omega}{|\boldsymbol{q}|} - \frac{\omega^2 - |\boldsymbol{q}|^2}{2|\boldsymbol{k}||\boldsymbol{q}|} \right) \\
                & \times \theta \left( |\boldsymbol{p}| - \omega \right) \theta \left( \omega + |\boldsymbol{k}| \right) \theta \left( |\boldsymbol{q}| - |\omega| \right) \theta \left( |\boldsymbol{p}| - \frac{|\boldsymbol{q}| + \omega}{2} \right) \theta \left( |\boldsymbol{k}| - \frac{|\boldsymbol{q}| - \omega}{2} \right) \\
                & \times e^{- \frac{|\boldsymbol{k}|}{\alpha \left( \tau \right)}} \left( \omega \left( |\boldsymbol{p}| - |\boldsymbol{k}| \right) - \omega^2 \right) \left( 1 - \left( \cos \theta_{\boldsymbol{pq}} \cos \theta_{\boldsymbol{kq}} + \cos \left( \phi_{\boldsymbol{p}} - \phi_{\boldsymbol{k}} \right) \sin \theta_{\boldsymbol{pq}} \sin \theta_{\boldsymbol{kq}} \right) \right) \\
                 = & \frac{1}{ \left( 2 \pi \right)^4} e^{- \frac{|\boldsymbol{p}|}{\alpha \left( \tau \right)}} A_1^2 \left( \tau \right) \int \mathrm{d}|\boldsymbol{k}| \mathrm{d} \phi_{\boldsymbol{k}} \mathrm{d}|\boldsymbol{q}| \mathrm{d} \phi_{\boldsymbol{q}} \mathrm{d} \omega e^{- \frac{|\boldsymbol{k}|}{\alpha \left( \tau \right)}} |\boldsymbol{k}| \left( \omega \left( |\boldsymbol{p}| - |\boldsymbol{k}| \right) - \omega^2 \right) \\
                & \times \theta \left( |\boldsymbol{p}| - \omega \right) \theta \left( \omega + |\boldsymbol{k}| \right) \theta \left( |\boldsymbol{q}| - |\omega| \right) \theta \left( |\boldsymbol{p}| - \frac{|\boldsymbol{q}| + \omega}{2} \right) \theta \left( |\boldsymbol{k}| - \frac{|\boldsymbol{q}| - \omega}{2} \right) \\
                & \times \left[ 1 - \left( \left( \frac{\omega}{|\boldsymbol{q}|} - \frac{\omega^2 - |\boldsymbol{q}|^2}{2|\boldsymbol{p}||\boldsymbol{q}|} \right) \left( \frac{\omega}{|\boldsymbol{q}|} + \frac{\omega^2 - |\boldsymbol{q}|^2}{2|\boldsymbol{k}||\boldsymbol{q}|} \right) \right. \right. \\
                & \left. \left. + \cos \left( \phi_{\boldsymbol{p}} - \phi_{\boldsymbol{k}} \right) \sqrt{1 - \left( \frac{\omega}{|\boldsymbol{q}|} - \frac{\omega^2 - |\boldsymbol{q}|^2}{2|\boldsymbol{p}||\boldsymbol{q}|} \right)^2} \sqrt{1 - \left( \frac{\omega}{|\boldsymbol{q}|} + \frac{\omega^2 - |\boldsymbol{q}|^2}{2|\boldsymbol{k}||\boldsymbol{q}|} \right)^2} \right) \right]. \label{226} \end{split} \end{align}
Finally, by translating all step functions into constraints on the upper and lower limits of integration, we arrive at
\begin{align} \begin{split} C \left[ f \right] = & \frac{1}{ \left( 2 \pi \right)^2} e^{- \frac{|\boldsymbol{p}|}{\alpha \left( \tau \right)}} A_1^2 \left( \tau \right) \left[
                \int_{\frac{|\boldsymbol{q}| - \omega}{2}}^{+ \infty}\mathrm{d}|\boldsymbol{k}|\int_{-|\boldsymbol{q}|}^{+|\boldsymbol{q}|} \mathrm{d} \omega \int_0^{|\boldsymbol{p}|}\mathrm{d}|\boldsymbol{q}| e^{- \frac{|\boldsymbol{k}|}{\alpha \left( \tau \right)}} |\boldsymbol{k}| \left( \omega \left( |\boldsymbol{p}| - |\boldsymbol{k}| \right) - \omega^2 \right) \right.\\
                &\left. \times \left( 1 - \left( \frac{\omega}{|\boldsymbol{q}|} - \frac{\omega^2 - |\boldsymbol{q}|^2}{2|\boldsymbol{p}||\boldsymbol{q}|} \right) \left( \frac{\omega}{|\boldsymbol{q}|} + \frac{\omega^2 - |\boldsymbol{q}|^2}{2|\boldsymbol{k}||\boldsymbol{q}|} \right) \right) \right.\\
                &\left. + \int_{\frac{|\boldsymbol{q}| - \omega}{2}}^{+ \infty}\mathrm{d}|\boldsymbol{k}| \int_{-|\boldsymbol{q}|}^{2|\boldsymbol{p}| - |\boldsymbol{q}|}\mathrm{d} \omega \int_{|\boldsymbol{p}|}^{+ \infty}\mathrm{d}|\boldsymbol{q}|     e^{- \frac{|\boldsymbol{k}|}{\alpha \left( \tau \right)}} |\boldsymbol{k}| \left( \omega \left( |\boldsymbol{p}| - |\boldsymbol{k}| \right) - \omega^2 \right) \right.\\
                &\left. \times \left( 1 - \left( \frac{\omega}{|\boldsymbol{q}|} - \frac{\omega^2 - |\boldsymbol{q}|^2}{2|\boldsymbol{p}||\boldsymbol{q}|} \right) \left( \frac{\omega}{|\boldsymbol{q}|} + \frac{\omega^2 - |\boldsymbol{q}|^2}{2|\boldsymbol{k}||\boldsymbol{q}|} \right) \right) \right] \\
                = & \frac{1}{6 \pi^2} \left[ e^{- \frac{|\boldsymbol{p}|}{\alpha \left( \tau \right)}} A_1^2 \left( \tau \right) |\boldsymbol{p}| \left( |\boldsymbol{p}| - 6 \alpha \left( \tau \right) \right) \left( |\boldsymbol{p}| - 2 \alpha \left( \tau \right) \right) \alpha^3 \left( \tau \right) \right] \\
                = & e^{- \frac{p^0}{\alpha \left( \tau \right)}} \left[ \frac{2 A_1^2 \left( \tau \right) \alpha^5 \left( \tau \right)} {\pi^2} p^0 - \frac{4 A_1^2 \left( \tau \right) \alpha^4 \left( \tau \right)}{3 \pi^2} \left( p^0 \right)^2 + \frac{A_1^2 \left( \tau \right) \alpha^3 \left( \tau \right)}{6 \pi^2} \left( p^0 \right)^3 \right]. \label{227} \end{split} \end{align}
 Remarkably, the collision integral\textemdash typically intractable in closed form\textemdash can be worked out \textit{analytically}. This is undoubtedly a nontrivial result, as even a slight increase in interaction complexity\textemdash such as the inclusion of angular dependence in the scattering cross section\textemdash would render the integral analytically intractable.
 After completing this nontrivial collision integral, we compare the powers of $p^0$ on both sides of the Boltzmann equation, which allows us to obtain
\begin{align} \begin{split} 
                A_0' \left( \tau \right) & = \frac{2 A_1^2 \left( \tau \right) \alpha^5 \left( \tau \right)}{\pi^2}, \\
                A_1' \left( \tau \right) + \frac{A_0 \left( \tau \right) \alpha' \left( \tau \right)}{\alpha^2 \left( \tau \right)} & = - \frac{4 A_1^2 \left( \tau \right) \alpha^4 \left( \tau \right)}{3 \pi^2}, \\
                \frac{A_1 \left( \tau \right) \alpha' \left( \tau \right)}{\alpha^2 \left( \tau \right)} & = \frac{A_1^2 \left( \tau \right) \alpha^3 \left( \tau \right)}{6 \pi^2}. \label{228} \end{split} \end{align}
Solving this system of ordinary differential equations directly is rather difficult. It should be noted that the particle number density  $n_0$ and the energy density $e_0$ are the conserved quantities in the homogeneous case (see also \cite{Hu:2024utr} for details) with their definitions 
\begin{align} \begin{split} n_0 = \int \frac{\mathrm{d}^3 \boldsymbol{p}}{ \left( 2 \pi \right)^3} f \left( \tau, p^0 \right), \qquad e_0 = \int \frac{\mathrm{d}^3 \boldsymbol{p}}{\left( 2 \pi \right)^3} p^0 f \left( \tau, p^0 \right). \label{229} \end{split} \end{align}
These two conserved quantities provide the initial conditions necessary for solving the equations. Combining Eq.\eqref{229} with Eq.\eqref{223}, $A_0 \left( \tau \right)$ and $A_1 \left( \tau \right)$ can be expressed as a combination of $n_0$, $e_0$, and $\alpha \left( \tau \right)$,
\begin{align} \begin{split} A_0 \left( \tau \right) = \frac{\pi^2 \left( -e_0 + 4 n_0 \alpha \left( \tau \right) \right)}{\alpha^4 \left( \tau \right)}, \quad A_1 \left( \tau \right) = \frac{\pi^2 \left( e_0 - 3 n_0 \alpha \left( \tau \right) \right)}{3 \alpha^5 \left( \tau \right)}. \label{230} \end{split} \end{align}
The detailed derivation of these expressions is provided in Appendix \ref{Determination}.

Substituting Eq.\eqref{230} back into the first equation in Eq.\eqref{228} yields the following solution:
\begin{align} \begin{split} \alpha \left( \tau \right) = \frac{e_0}{3 n_0} + \eta e^{- \frac{n_0 \tau}{6}}, \label{231} \end{split} \end{align}
\noindent where $\eta$ is a constant independent of $p^0$ and $\tau$ acting as one free initial parameter. Note Eqs.\eqref{230} and \eqref{231} are also consistent with the other two equations in Eq.\eqref{228} (see Appendix \ref{Determination}). We also verify that our solutions reproduce the result given in \cite{Hu:2024utr} in the case of hard-sphere interaction. For comparison with \cite{Bazow:2015dha,Bazow:2016oky}, we specify the initial condition by choosing the value of $\eta$ as $\eta = -\frac{1}{4}$, then the resulting expression reproduces the solution given in Ref.\cite{Bazow:2015dha,Bazow:2016oky} in Minkowski spacetime. A similar discussion on how the choice of initial conditions affects the final physical results can be found in Ref.\cite{Hu:2024utr}. It is also worthwhile to note that the present calculations can be readily extended to isotropically expanding FLRW spacetime \cite{weinberg}.

\section{Lie algebra of the relativistic Boltzmann equation} 
\label{Extended Symmetry Generators for the Relativistic Boltzmann Equation and Applications}
To avoid ambiguity, we would like to kindly remind the reader that in this section, $f$ refers to the general distribution function discussed in the Lie algebra analysis and is not restricted to the form of Eq. (21) or other exact analytical solutions of the Boltzmann equation presented in Sec. \ref{Lie Groups, Lie Algebras, and Exact Analytical Solutions for the Boltzmann Equation}.
\subsection{Symmetry transformation and Lie algebra}
\label{Symmetry Generators and Their Associated Lie Algebra}
Let us first consider a general group transformation
\begin{align} \begin{split} & f \to f' = \mathcal{A} \left( x^{\mu}, p^{\mu}, f, \theta \right), \\
                & x^{\mu} \to x'^{\mu} = \mathcal{B}^{\mu} \left( x^{\mu}, p^{\mu}, f, \theta \right), \\
                & p^{\mu} \to p'^{\mu} = \mathcal{C}^{\mu} \left( x^{\mu}, p^{\mu}, f, \theta \right), \label{31} \end{split} \end{align}
\noindent with initial conditions
\begin{align} \begin{split} & \mathcal{A} \left( x^{\mu}, p^{\mu}, f, 0 \right) = f, \\
                & \mathcal{B}^{\mu} \left( x^{\mu}, p^{\mu}, f, 0 \right) = x^{\mu}, \\
                & \mathcal{C}^{\mu} \left( x^{\mu}, p^{\mu}, f, 0 \right) = p^{\mu}, \label{32} \end{split} \end{align}
\noindent where $\theta$ is the group parameter. Expanding the functions $\mathcal{A}$, $\mathcal{B}^{\mu}$, and $\mathcal{C}^{\mu}$ in a Taylor series around $\theta = 0$, while taking into account the initial conditions \eqref{32}, one can obtain infinitesimal transformations 
\begin{align} \begin{split} & f' \approx f + \theta \Omega \left( x^{\mu}, p^{\mu}, f \right), \\
                & x'^{\mu} \approx x^{\mu} + \theta X^{\mu} \left( x^{\mu}, p^{\mu}, f \right), \\
                & p'^{\mu} \approx p^{\mu} + \theta P^{\mu} \left( x^{\mu}, p^{\mu}, f \right), \label{33} \end{split} \end{align}
\noindent where
\begin{align} \begin{split} & \Omega \left( x^{\mu}, p^{\mu}, f \right) = \frac{\partial \mathcal{A} \left( x^{\mu}, p^{\mu}, f, \theta \right)}{\partial \theta} \big{|}_{\theta = 0}, \\
                & X^{\mu} \left( x^{\mu}, p^{\mu}, f \right) = \frac{\partial \mathcal{B}^{\mu} \left( x^{\mu}, p^{\mu}, f, \theta \right)}{\partial \theta} \big{|}_{\theta = 0}, \\
                & P^{\mu} \left( x^{\mu}, p^{\mu}, f \right) = \frac{\partial \mathcal{C}^{\mu} \left( x^{\mu}, p^{\mu}, f, \theta \right)}{\partial \theta} \big{|}_{\theta = 0}. \label{34} \end{split} \end{align}
Subsequently, we are able to obtain the generator of the group
\begin{align} \begin{split} \hat{L} = \Omega \left( x^{\mu}, p^{\mu}, f \right) \frac{\partial}{\partial f} + X^{\mu} \left( x^{\mu}, p^{\mu}, f \right) \frac{\partial}{\partial x^{\mu}} + P^{\mu} \left( x^{\mu}, p^{\mu}, f \right) \frac{\partial}{\partial p^{\mu}}. \label{35} \end{split} \end{align}
If we have infinitesimal transformations \eqref{33} or generator \eqref{35}, the transformation \eqref{31} can be defined using the following Lie equations:
\begin{align} \begin{split} & \frac{\mathrm{d} \mathcal{A}}{\mathrm{d} \theta} = \Omega \left( \mathcal{A}, \mathcal{B}^{\mu}, \mathcal{C}^{\mu} \right), \qquad \mathcal{A} \big{|}_{\theta = 0} = f, \\
                & \frac{\mathrm{d} \mathcal{B}^{\mu}}{\mathrm{d} \theta} = X^{\mu} \left( \mathcal{A}, \mathcal{B}^{\mu}, \mathcal{C}^{\mu} \right), \qquad \mathcal{B}^{\mu} \big{|}_{\theta = 0} = x^{\mu}, \\
                & \frac{\mathrm{d} \mathcal{C}^{\mu}}{\mathrm{d} \theta} = P^{\mu} \left( \mathcal{A}, \mathcal{B}^{\mu}, \mathcal{C}^{\mu} \right), \qquad \mathcal{C}^{\mu} \big{|}_{\theta = 0} = p^{\mu}. \label{36} \end{split} \end{align}

We aim to obtain certain symmetry transformations, and then use them to map any solution of the Boltzmann equation back onto another solution of the same equation.  That is, let Eq.\eqref{31} be a symmetry transformation of Eq.\eqref{21}, and let the function
\begin{align} f = \Phi \left( x^{\mu}, p^{\mu} \right)  \end{align}
\noindent be a solution of Eq.\eqref{21}. Since Eq.\eqref{31} defines a symmetry transformation, the above solution can equivalently be expressed in terms of the transformed variables as
\begin{align} f' = \Phi \left( x'^{\mu}, p'^{\mu} \right),  \end{align}
\noindent where $f'$ satisfies the Boltzmann equation formulated with respect to the transformed coordinates and momenta $\left( x'^{\mu}, p'^{\mu} \right)$. Substituting the transformation relations from Eq.~\eqref{31} into the expression above, we finally obtain 
\begin{align} \begin{split} \mathcal{A} \left( x^{\mu}, p^{\mu}, f, \theta \right) = \Phi \left( \mathcal{B}^{\mu} \left( x^{\mu}, p^{\mu}, f, \theta \right), \mathcal{C}^{\mu} \left( x^{\mu}, p^{\mu}, f, \theta \right) \right). \label{37} \end{split} \end{align}
If Eq.\eqref{37} with respect to $f$ is solved, we can then obtain a new solution of Eq.\eqref{21}. Because the Boltzmann equation is a nonlinear integro-differential equation, its determining equations for Lie symmetries become highly nonlinear and analytically intractable. This difficulty motivates a shift in perspective: rather than solving the full system directly, we seek the corresponding Lie algebra following a problem-oriented and physics-informed approach. With these preparations in place, the following subsection will demonstrate how to achieve this.

\subsection{Solving for the Lie algebra}
\label{Lie Group Structure of the Relativistic Boltzmann Equation}
The relativistic Boltzmann equation \eqref{21} is formally invariant under Poincaré transformations by construction. Therefore, it is natural to show that the following  generators
\begin{align} \begin{split} & \hat{L}^{\left( 0 \right)}_{ij} = \left( x_i \frac{\partial}{\partial x^j} - x_j \frac{\partial}{\partial x^i} \right) + \left( p_i \frac{\partial}{\partial p^j} - p_j \frac{\partial}{\partial p^i} \right), \\
                            & \hat{L}^{\left( 1 \right)}_i = \frac{\partial}{\partial x^i}, \\
                            & \hat{L}^{\left( 2 \right)}_i = \left( x_0 \frac{\partial}{\partial x^i} - x_i \frac{\partial}{\partial x^0} \right) + \left( p_0 \frac{\partial}{\partial p^i} - p_i \frac{\partial}{\partial p^0} \right), \\
                            & \hat{L}^{\left( 3 \right)} = \frac{\partial}{\partial x^0}. 
                            \label{38} \end{split} \end{align}
\noindent These generators form a basic and closed Lie algebra.  However, they still differ from the conventional Poincaré algebra. Specifically, the generators $\hat{L}^{\left( 0 \right)}_{ij}$ and $\hat{L}^{\left( 2 \right)}_i$ generate rotations and boost transformations in phase space, whereas spacetime translations remain confined to coordinate space. Since this algebra is trivially closed, no new elements can be generated through Lie brackets. This also implies that only by incorporating new physical insights can we obtain additional generators to construct a larger Lie algebra.

Recalling that in Sec.\ref{known}, we have presented several analytical solutions of the Boltzmann equation, which could offer rich physical insights. If a particular solution exhibits a certain symmetry, that symmetry should be consistent with the structure of the Boltzmann equation. In other words, by examining the symmetries manifested in these physical solutions, we can uncover the underlying symmetries of the Boltzmann equation itself. For example, the solution described by Eq.\eqref{210} is precisely such a candidate that may provide valuable insight, as it respects the following scaling symmetry: If we perform the transformation 
\begin{equation} t \rightarrow \lambda t, \quad p^0 \rightarrow p^0 \lambda^{- \tilde{b}}, \label{39} \end{equation}
then the distribution function transforms accordingly as $f\rightarrow \lambda^{\tilde{a}} f$. It is important to note that the scaling behaviors described above are not universally valid. The self-similar scaling in Eq.\eqref{39} reflects the property of the NTFP within specific regimes; see \cite{Berges:2008wm,Berges:2008mr,PineiroOrioli:2015cpb} and also Sec.\ref{known} for more details. Inspired by the scaling symmetry of the NTFP, it is natural to ask whether there is a similar scaling symmetry respected by the Boltzmann equation. It is easy to draw a conclusion that the following scaling transformation generator:
\begin{align} \begin{split} \hat{L}^{\left( 4 \right)} = x^{\mu} \frac{\partial}{\partial x^{\mu}} - f \frac{\partial}{\partial f} \label{310} \end{split} \end{align}
\noindent together with the resulting scale transformation
\begin{align} x^\mu \rightarrow x^{\prime \mu} = e^\theta x^\mu, \quad p^\mu \rightarrow p^{\prime\mu} = p^\mu, \quad f \rightarrow f^\prime = e^{- \theta} f \label{311} \end{align}
is the qualified one. It should be emphasized, however, that the scaling transformation generated by $\hat{L}^{(4)}$ is an exact symmetry only for the classical Boltzmann equation; the introduction of quantum statistical factors $(1 \pm af)$ in the collision kernel explicitly breaks this symmetry, as the constant `1' does not scale with the distribution function $f$.

It is evident that under the scaling transformation generated by $\hat{L}^{\left( 4 \right)}$, the momentum remains invariant, in contrast to the scaling transformation in Eq.\eqref{39}. If the cross section transforms as $\sigma \to \lambda^{2 \left( r - 1 \right)} \sigma$ under $p^{\mu} \to \lambda p^{\mu}$, its specific form coincides with Eq.\eqref{25} when $r = l$, which in turn allows us to identify another independent generator and its corresponding transformation
\begin{align} \begin{split} & \hat{L}^{\left( 5 \right)} = v x^{\mu} \frac{\partial}{\partial x^{\mu}} - \frac{1}{2r + 1} p^{\mu} \frac{\partial}{\partial p^{\mu}} + \left( 1 - v \right) f \frac{\partial}{\partial f}, \\
                 & x^\mu \rightarrow x^{\prime\mu} = e^{v \theta} x^\mu, \quad p^\mu \rightarrow p^{\prime \mu} = e^{- \frac{1}{2r + 1} \theta} p^\mu, \quad f \rightarrow   f^\prime = e^{\left( 1 - v \right) \theta} f,
                 \label{313} \end{split} \end{align}
where $v$ is an arbitrary real constant that parametrizes this family of scaling transformations, representing the relative weight between the scaling of spacetime coordinates and the distribution function. Although $\hat{L}^{\left( 4 \right)}$ and $\hat{L}^{\left( 5 \right)}$ both appear to generate scaling transformations, they are distinct: $\hat{L}^{\left( 5 \right)}$ cannot be reduced to $\hat{L}^{\left( 4 \right)}$. As in the case of $\hat{L}^{\left( 4 \right)}$, the inclusion of  $\hat{L}^{\left( 5 \right)}$ within the algebra generated by $\hat{L}^{\left( 0 \right)}$ – $\hat{L}^{\left( 4 \right)}$ results in a closed structure.

\subsection{Discussion on the Lie algebra}
\label{Properties of Extended Symmetry Generators}
In the previous subsection, we derived the Lie algebra of invariant transformations for the relativistic Boltzmann equation. Here, we summarize these results and present a physical discussion. The complete set of generators forming the Lie algebra is presented below
\begin{align} \begin{split} & \hat{L}^{\left( 0 \right)}_{ij} = \left( x_i \frac{\partial}{\partial x^j} - x_j \frac{\partial}{\partial x^i} \right) + \left( p_i \frac{\partial}{\partial p^j} - p_j \frac{\partial}{\partial p^i} \right), \\
                & \hat{L}^{\left( 1 \right)}_i = \frac{\partial}{\partial x^i}, \\
                & \hat{L}^{\left( 2 \right)}_i = \left( x_0 \frac{\partial}{\partial x^i} - x_i \frac{\partial}{\partial x^0} \right) + \left( p_0 \frac{\partial}{\partial p^i} - p_i \frac{\partial}{\partial p^0} \right), \\
                & \hat{L}^{\left( 3 \right)} = \frac{\partial}{\partial x^0}, \\
                & \hat{L}^{\left( 4 \right)} = x^{\mu} \frac{\partial}{\partial x^{\mu}} - f \frac{\partial}{\partial f}, \\
                & \hat{L}^{\left( 5 \right)} = v x^{\mu} \frac{\partial}{\partial x^{\mu}} - \frac{1}{2r + 1} p^{\mu} \frac{\partial}{\partial p^{\mu}} + \left( 1 - v \right) f \frac{\partial}{\partial f}. \label{314} \end{split} \end{align}
The range of applicability of the transformations generated by these generators varies. The generators $\hat{L}^{\left( 0 \right)}$ – $\hat{L}^{\left( 3 \right)}$ form a Poincaré algebra, which applies to Eq.\eqref{21}, regardless of whether quantum statistics are included or whether the particle mass is zero. In contrast, $\hat{L}^{\left( 4 \right)}$ is applicable exclusively to classical systems and remains valid independently of the particle mass. On the other hand, $\hat{L}^{\left( 5 \right)}$ is valid only for massless classical systems, where the cross section exhibits scaling behavior $\sigma \to \lambda^{2 \left( r - 1 \right)} \sigma$ under the transformation $p^{\mu} \to \lambda p^{\mu}$.\footnote{There is one exception concerning $\hat{L}^{\left( 5 \right)}$: when $v = 1$, it remains valid for the Boltzmann equation with quantum statistics.} Furthermore, it can be shown that the entire set of generators given by Eq.\eqref{314} is closed under the Lie bracket, as summarized in Table \ref{tab1}. To the best of our knowledge, this is the first time that a closed Lie algebra of invariant transformations for the relativistic Boltzmann equation has been presented.
\begin{table}[H]
\centering
\begin{tabular}{|c|c|c|c|c|c|c|}
\hline
\cline{1-7}
\diagbox[width=4em,height=2.5em]{\raisebox{0.25em}{$\hat{L}_{\beta}$}}{\raisebox{-0.25em}{$\hat{L}_{\alpha}$}} & $\hat{L}^{\left( 0 \right)}_{ij}$ & $\hat{L}^{\left( 1 \right)}_i$ & $\hat{L}^{\left( 2 \right)}_i$ & $\hat{L}^{\left( 3 \right)}$ & $\hat{L}^{\left( 4 \right)}$ & $\hat{L}^{\left( 5 \right)}$ \\ 
\hline
\rule{0pt}{3ex} $\hat{L}^{\left( 0 \right)}_{mn}$ & $g_{jm} \hat{L}^{\left( 0 \right)}_{in} + g_{in} \hat{L}^{\left( 0 \right)}_{jm} + g_{jn} \hat{L}^{\left( 0 \right)}_{mi} + g_{im} \hat{L}^{\left( 0 \right)}_{nj}$ & $\backslash$ & $\backslash$ & $\backslash$ & $\backslash$ & $\backslash$ \\
\hline
\rule{0pt}{3ex} $\hat{L}^{\left( 1 \right)}_m$ & $g_{jm} \hat{L}^{\left( 1 \right)}_i - g_{im} \hat{L}^{\left( 1 \right)}_j$ & $0$ & $\backslash$ & $\backslash$ & $\backslash$ & $\backslash$ \\
\hline
\rule{0pt}{3ex} $\hat{L}^{\left( 2 \right)}_m$ & $g_{jm} \hat{L}^{\left( 2 \right)}_i - g_{im} \hat{L}^{\left( 2 \right)}_j$ & $-g_{im} \hat{L}^{\left( 3 \right)}$ & $\hat{L}^{\left( 0 \right)}_{mi}$ & $\backslash$ & $\backslash$ & $\backslash$ \\
\hline
\rule{0pt}{3ex} $\hat{L}^{\left( 3 \right)}$ & $0$ & $0$ & $- \hat{L}^{\left( 1 \right)}_i$ & $0$ & $\backslash$ & $\backslash$ \\
\hline
\rule{0pt}{3ex} $\hat{L}^{\left( 4 \right)}$ & $0$ & $\hat{L}^{\left( 1 \right)}_i$ & $0$ & $\hat{L}^{\left( 3 \right)}$ & $0$ & $\backslash$ \\
\hline
\rule{0pt}{3ex} $\hat{L}^{\left( 5 \right)}$ & $0$ & $v \hat{L}^{\left( 1 \right)}_i$ & $0$ & $v \hat{L}^{\left( 3 \right)}$ & $0$ & $0$ \\
\hline
\end{tabular}
\caption{The commutators between the Lie algebra generators are presented as $[\hat{L}_{\alpha}, \hat{L}_{\beta}]$. The backslash symbols ($\backslash$) denote redundant entries that are omitted due to the antisymmetry of the Lie bracket, i.e., $[\hat{L}_{\alpha}, \hat{L}_{\beta}] = -[\hat{L}_{\beta}, \hat{L}_{\alpha}]$. For those entries not explicitly listed, they can be obtained from the corresponding existing entries by simply adding a minus sign and exchanging the indices.}
\label{tab1} \end{table}

The given Lie algebra formed by $L^{\left( 0 \right)}$ to $L^{\left( 5 \right)}$ has a one-to-one correspondence with its nonrelativistic counterpart \cite{Bobylev:1996, bobylev1993} except for the following element $\hat{L}^{\left( 6 \right)}$. Specifically, the nonrelativistic Boltzmann equation admits an additional generator, $\hat{L}^{\left( 6 \right)}$, associated with a nonlinear transformation,
\begin{align} \begin{split} & \hat{L}^{\left( 6 \right)} = t^2 \frac{\partial}{\partial t} + t \boldsymbol{x} \frac{\partial}{\partial \boldsymbol{x}} + \left( \boldsymbol{x} - \boldsymbol{v} t \right) \frac{\partial}{\partial \boldsymbol{v}}, \\
                & f^{\left( 6 \right)}_{\theta} \left( t, \boldsymbol{x}, \boldsymbol{v} \right) = f \left( \frac{t}{1 - \theta t}, \frac{\boldsymbol{x}}{1 - \theta t}, \boldsymbol{v} + \theta \left( \boldsymbol{x} - \boldsymbol{v} t \right) \right), \label{315} \end{split} \end{align}
where $\boldsymbol{v}$ is the particle velocity. According to Refs.\cite{Bobylev:1996, bobylev1993}, the above generator is valid only when the system is under the power law potential $U \left( r \right) \propto r^{-m}$ with $m = 2$; see \cite{Bobylev:1996, bobylev1993} for more details.  However, due to the stringent requirements of relativistic covariance and the on-shell condition for particles, no generator analogous to $\hat{L}^{\left( 6 \right)}$ can be constructed in the relativistic regime.

\section{Conclusion and outlook}
\label{Conclusion and outlook}
In this work, we provide a novel and efficient approach to constructing the relativistic BKW solution of the Boltzmann equation. We first introduce a class of ansatz functions for the single-particle distribution function, and then demonstrate that within this specific functional space, only two forms yield exact solutions: the equilibrium distribution and the BKW-type solution. This analysis not only gives a streamlined derivation compared to conventional moment methods but also offers insight into the uniqueness of these solutions.

Furthermore, guided by physical intuition and the structure of known analytical solutions, particularly the NTFP, we have systematically derived the closed Lie algebra of invariant transformations admitted by the relativistic Boltzmann equation. To the best of our knowledge, this is the first time such a closed and comprehensive symmetry algebra has been constructed in the context of relativistic kinetic theory. The resulting algebra, spanned by generators  $\hat{L}^{\left( 0 \right)}$ to $\hat{L}^{\left( 5 \right)}$, encompasses Poincaré symmetry ($\hat{L}^{\left( 0 \right)}$ – $\hat{L}^{\left( 3 \right)}$), spacetime scaling ($\hat{L}^{\left( 4 \right)}$), and a more general momentum-rescaling transformation ($\hat{L}^{\left( 5 \right)}$) tied to the specific interactions of scaling cross section.

There are several avenues for future work that can be pursued based on the current findings. First, we can explore whether there is a better approach to provide a first-principle determination of the Lie algebra, which might help us exhaustively enumerate all physically relevant invariant Lie algebras. Second, the properties of the linearized collision operator around the BKW solution are also worth investigating. How does it differ from the linearized collision operator around the equilibrium state? What distinct behaviors might emerge in the linear response around the BKW solution? As an exact analytical solution of the Boltzmann equation, the BKW solution holds the potential to offer insights into how to probe the linear response behavior of nonequilibrium states within kinetic theory. Additionally, as another self-similar solution of the relativistic Boltzmann equation, the connection between the BKW solution and the NTFP has yet to be explored. This presents an important direction for future research that could yield significant new insights.

\vspace{1cm}
\section*{ACKNOWLEDGMENTS}
We express our gratitude for the valuable discussions with Qiuze Sun, Tianzhe Zhou, Xiaotian Ma, Haiyang Shao, Shuzhe Shi, Qi Chen, Wan Wu, Luyao Li, Baiting Tian, Puyuan Bai, Weiyao Ke, Yi Yin, and others. This work was financially supported by the National Natural Science Foundation of China under Grants No. 12035006 and No. 12505149. 

\begin{appendix}

\section{EVALUATION OF THE COLLISION INTEGRAL AND KINEMATIC CONSTRAINTS}
\label{AppendixCollision}
{
In this appendix, we provide the detailed mathematical steps for evaluating the Dirac delta functions in Eq.\,\eqref{212} and deriving the associated kinematic boundaries requested in the main text. 

From the arguments of the delta functions in Eq.\,\eqref{212}, the positivity of the final-state momentum magnitudes, $|\boldsymbol{p}_f| \ge 0$ and $|\boldsymbol{k}_f| \ge 0$, imposes the conditions $|\boldsymbol{p}| - \omega \ge 0$ and $\omega + |\boldsymbol{k}| \ge 0$, respectively. These physical requirements are formally enforced by introducing the Heaviside step functions $\theta(|\boldsymbol{p}| - \omega)$ and $\theta(\omega + |\boldsymbol{k}|)$.

To express the delta functions in terms of the scattering angles, we employ the standard identity $\delta(h(x)) = \sum_i |h'(x_i)|^{-1} \delta(x - x_i)$, where $x_i$ are the simple roots of $h(x) = 0$. Using the geometric relations, 
\begin{align}
    |\boldsymbol{p}_f| &= \sqrt{|\boldsymbol{p} - \boldsymbol{q}|^2} = \sqrt{\boldsymbol{p}^2 + \boldsymbol{q}^2 - 2|\boldsymbol{p}||\boldsymbol{q}| \cos \theta_{\boldsymbol{pq}}}, \label{cospq} \\
    |\boldsymbol{k}_f| &= \sqrt{|\boldsymbol{k} + \boldsymbol{q}|^2} = \sqrt{\boldsymbol{k}^2 + \boldsymbol{q}^2 + 2|\boldsymbol{k}||\boldsymbol{q}| \cos \theta_{\boldsymbol{kq}}}, \label{coskq}
\end{align}
we define the auxiliary functions $M(\cos \theta_{\boldsymbol{pq}})$ and $N(\cos \theta_{\boldsymbol{kq}})$ as
\begin{align}
    M(\cos \theta_{\boldsymbol{pq}}) &\equiv|\boldsymbol{p}_f| - \left( |\boldsymbol{p}| - \omega \right)= \sqrt{\boldsymbol{p}^2 + \boldsymbol{q}^2 - 2|\boldsymbol{p}||\boldsymbol{q}| \cos \theta_{\boldsymbol{pq}}} - (|\boldsymbol{p}| - \omega), \label{M function} \\
    N(\cos \theta_{\boldsymbol{kq}}) &\equiv  |\boldsymbol{k}_f| - \left( \omega + |\boldsymbol{k}| \right)=\sqrt{\boldsymbol{k}^2 + \boldsymbol{q}^2 + 2|\boldsymbol{k}||\boldsymbol{q}| \cos \theta_{\boldsymbol{kq}}} - (\omega + |\boldsymbol{k}|). \label{N function}
\end{align}
Equating these functions to zero yields the unique roots for the angular variables:
\begin{align}
    \cos \theta_{\boldsymbol{pq} 0} = \frac{\omega}{|\boldsymbol{q}|} - \frac{\omega^2 - |\boldsymbol{q}|^2}{2|\boldsymbol{p}||\boldsymbol{q}|}, \qquad 
    \cos \theta_{\boldsymbol{kq} 0} = \frac{\omega}{|\boldsymbol{q}|} + \frac{\omega^2 - |\boldsymbol{q}|^2}{2|\boldsymbol{k}||\boldsymbol{q}|}.
\end{align}
The absolute values of their derivatives evaluated at these roots are, respectively, given by
\begin{align}
    |M'(\cos \theta_{\boldsymbol{pq} 0})| = \frac{|\boldsymbol{p}||\boldsymbol{q}|}{|\boldsymbol{p}_f|}, \qquad 
    |N'(\cos \theta_{\boldsymbol{kq} 0})| = \frac{|\boldsymbol{k}||\boldsymbol{q}|}{|\boldsymbol{k}_f|}.
\end{align}
Consequently, the delta functions in Eq.\,\eqref{212} transform as
\begin{align} \begin{split} \delta \left( |\boldsymbol{p}_f| - \left( |\boldsymbol{p}| - \omega \right) \right) & = \delta \left( M \left( \cos \theta_{\boldsymbol{pq}} \right) \right) \\  
                & = \frac{\delta \left( \cos \theta_{\boldsymbol{pq}} - \cos \theta_{\boldsymbol{pq} 0} \right)}{|M^\prime \left( \cos \theta_{\boldsymbol{pq} 0} \right)|} \theta \left( |\boldsymbol{p}| - \omega \right) \\
                & = \frac{|\boldsymbol{p}_f|}{|\boldsymbol{p}||\boldsymbol{q}|} \delta \left( \cos \theta_{\boldsymbol{pq}} - \frac{\omega}{|\boldsymbol{q}|} + \frac{\omega^2 - |\boldsymbol{q}|^2}{2|\boldsymbol{p}||\boldsymbol{q}|} \right) \theta \left( |\boldsymbol{p}| - \omega \right), \end{split} \end{align}
and
\begin{align} \begin{split} \delta \left( |\boldsymbol{k}_f| - \left( \omega + |\boldsymbol{k}| \right)\right) & = \delta \left( N \left( \cos \theta_{\boldsymbol{kq}} \right) \right) \\  
                & = \frac{\delta \left( \cos \theta_{\boldsymbol{kq}} - \cos \theta_{\boldsymbol{kq} 0} \right)}{|N^\prime \left( \cos \theta_{\boldsymbol{kq} 0} \right)|} \theta \left( \omega + |\boldsymbol{k}| \right) \\
                & = \frac{|\boldsymbol{k}_f|}{|\boldsymbol{k}||\boldsymbol{q}|} \delta \left( \cos \theta_{\boldsymbol{kq}} - \frac{\omega}{|\boldsymbol{q}|} - \frac{\omega^2 - |\boldsymbol{q}|^2}{2|\boldsymbol{k}||\boldsymbol{q}|} \right) \theta \left( \omega + |\boldsymbol{k}| \right).  \end{split} \end{align}
Thus, we obtain
\begin{align}
    &\delta \left( |\boldsymbol{p}_f| - \left( |\boldsymbol{p}| - \omega \right) \right) \delta \left( |\boldsymbol{k}_f| - \left( \omega + |\boldsymbol{k}| \right)\right) \nonumber \\
    &= \frac{|\boldsymbol{p}_f| |\boldsymbol{k}_f|}{|\boldsymbol{p}| |\boldsymbol{k}| |\boldsymbol{q}|^2} 
    \delta\left(\cos \theta_{\boldsymbol{pq}} - \frac{\omega}{|\boldsymbol{q}|} + \frac{\omega^2 - |\boldsymbol{q}|^2}{2|\boldsymbol{p}||\boldsymbol{q}|}\right) 
    \delta\left(\cos \theta_{\boldsymbol{kq}} - \frac{\omega}{|\boldsymbol{q}|} - \frac{\omega^2 - |\boldsymbol{q}|^2}{2|\boldsymbol{k}||\boldsymbol{q}|}\right)
    \theta(|\boldsymbol{p}| - \omega) \theta(\omega + |\boldsymbol{k}|).
\end{align}

Finally, we establish the kinematic bounds governing the scattering process. The spacelike nature of the momentum transfer dictates that $\omega^2 - |\boldsymbol{q}|^2 \le 0$, which yields the condition $\theta(|\boldsymbol{q}| - |\omega|)$. Furthermore, the geometric requirement that the cosines of the scattering angles must fall strictly within the physical domain $[-1, 1]$ introduces the following inequalities:
\begin{align}
    -1 \le \cos \theta_{\boldsymbol{pq}} \le 1 \quad &\Rightarrow \quad -1 \le \frac{\omega}{|\boldsymbol{q}|} - \frac{\omega^2 - |\boldsymbol{q}|^2}{2|\boldsymbol{p}||\boldsymbol{q}|} \le 1 \quad \Rightarrow \quad |\boldsymbol{p}| \ge \frac{|\boldsymbol{q}| + \omega}{2}, \\
    -1 \le \cos \theta_{\boldsymbol{kq}} \le 1 \quad &\Rightarrow \quad -1 \le \frac{\omega}{|\boldsymbol{q}|} + \frac{\omega^2 - |\boldsymbol{q}|^2}{2|\boldsymbol{k}||\boldsymbol{q}|} \le 1 \quad \Rightarrow \quad |\boldsymbol{k}| \ge \frac{|\boldsymbol{q}| - \omega}{2}.
\end{align}
Combining these geometric and kinematic requirements yields the piecewise step functions encapsulated in Eq.\eqref{219}.
}

\section{TRIAL FUNCTION FOR \texorpdfstring{$n > 1$}{n > 1}}
\label{Trial function}
We use $f_n \left( \tau, p^0 \right)$ to denote the single-particle distribution function given in Eq.\eqref{222}, whose highest power of $p^0$ is $n$. Then we have
\begin{align} \begin{split} & f_n \left( \tau, p_f^0 \right) f_n \left( \tau, k_f^0 \right) - f_n \left( \tau, p^0 \right) f_n \left( \tau, k^0 \right) \\
                = & f_{n - 1} \left( \tau, p_f^0 \right) f_{n - 1} \left( \tau, k_f^0 \right) - f_{n - 1} \left( \tau, p^0 \right) f_{n - 1} \left( \tau, k^0 \right) \\
                & + A_n \left( \tau \right) \left[ \left( p_f^0 \right)^n f_{n - 1} \left( \tau, k_f^0 \right) + \left( k_f^0 \right)^n f_{n - 1} \left( \tau, p_f^0 \right) - \left( p_f^0 \right)^n f_{n - 1} \left( \tau, k_f^0 \right) - \left( p_f^0 \right)^n f_{n - 1} \left( \tau, k_f^0 \right) \right] \\ 
                & + A_n^2 \left( \tau \right) \left[ \left( p_f^0 \right)^n \left( k_f^0 \right)^n - \left( p^0 \right)^n \left( k^0 \right)^n \right] \\
                = & f_{n - 1} \left( \tau, p^0 - \omega \right) f_{n - 1} \left( \tau, k^0 + \omega \right) - f_{n - 1} \left( \tau, p^0 \right) f_{n - 1} \left( \tau, k^0 \right) \\
                & + A_n \left( \tau \right) \left[ \left( p^0 - \omega \right)^n f_{n - 1} \left( \tau, k^0 + \omega \right) + \left( k^0 + \omega \right)^n f_{n - 1} \left( \tau, p^0 - \omega \right) \right. \\
                & \left. - \left( p^0 \right)^n f_{n - 1} \left( \tau, k^0 \right) - \left( k^0 \right)^n f_{n - 1} \left( \tau, p^0 \right) \right]  + A_n^2 \left( \tau \right) \left[ \left( p^0 - \omega \right)^n \left( k^0 + \omega \right)^n - \left( p^0 \right)^n \left( k^0 \right)^n \right], \label{A1} \end{split} \end{align}
where the above expression will be utilized to calculate the collision kernel. When $n = 2$,  the trial function takes the following form
\begin{align} \begin{split} f_2 \left( \tau, p^0 \right) = e^{- \frac{p^0}{\alpha \left( \tau \right)}} \left( A_0 \left( \tau \right) + A_1 \left( \tau \right) p^0 + A_2 \left( \tau \right) \left( p^0 \right)^2 \right), \label{A2} \end{split} \end{align}

\noindent and the collision term in Eq.\eqref{221} becomes 

\begin{align} \begin{split} C \left[ f_2 \right] = & C \left[ f_1 \right] + \frac{1}{\left( 2 \pi \right)^2} e^{- \frac{|\boldsymbol{p}|}{\alpha \left( \tau \right)}} \\
                & \times \left[ \int_{\frac{|\boldsymbol{q}| - \omega}{2}}^{+ \infty}\mathrm{d}|\boldsymbol{k}|\int_{-|\boldsymbol{q}|}^{+|\boldsymbol{q}|} \mathrm{d} \omega \int_0^{|\boldsymbol{p}|}\mathrm{d}|\boldsymbol{q}| e^{- \frac{|\boldsymbol{k}|}{\alpha \left( \tau \right)}} |\boldsymbol{k}| T \left( \tau, \boldsymbol{p}, \boldsymbol{k}, \omega \right) \left( 1 - \left( \frac{\omega}{|\boldsymbol{q}|} - \frac{\omega^2 - |\boldsymbol{q}|^2}{2|\boldsymbol{p}||\boldsymbol{q}|} \right) \left( \frac{\omega}{|\boldsymbol{q}|} + \frac{\omega^2 - |\boldsymbol{q}|^2}{2|\boldsymbol{k}||\boldsymbol{q}|} \right) \right) \right. \\
                & \left. + \int_{\frac{|\boldsymbol{q}| - \omega}{2}}^{+ \infty}\mathrm{d}|\boldsymbol{k}| \int_{-|\boldsymbol{q}|}^{2|\boldsymbol{p}| - |\boldsymbol{q}|}\mathrm{d} \omega \int_{|\boldsymbol{p}|}^{+ \infty}\mathrm{d}|\boldsymbol{q}| e^{- \frac{|\boldsymbol{k}|}{\alpha \left( \tau \right)}} |\boldsymbol{k}| T \left( \tau, \boldsymbol{p}, \boldsymbol{k}, \omega \right) \left( 1 - \left( \frac{\omega}{|\boldsymbol{q}|} - \frac{\omega^2 - |\boldsymbol{q}|^2}{2|\boldsymbol{p}||\boldsymbol{q}|} \right) \left( \frac{\omega}{|\boldsymbol{q}|} + \frac{\omega^2 - |\boldsymbol{q}|^2}{2|\boldsymbol{k}||\boldsymbol{q}|} \right) \right) \right] \\
                = & e^{- \frac{p^0}{\alpha \left( \tau \right)}} \left[ R_1 \left( \tau \right) p^0 + R_2 \left( \tau \right) \left( p^0 \right)^2 + R_3 \left( \tau \right) \left( p^0 \right)^3 + R_4 \left( \tau \right) \left( p^0 \right)^4 + R_5 \left( \tau \right) \left( p^0 \right)^5 \right], \label{A3} \end{split} \end{align}

\noindent where 
\begin{align} \begin{split} T \left( \tau, \boldsymbol{p}, \boldsymbol{k}, \omega \right) \equiv & A_2 \left( \tau \right) \left[ \left( |\boldsymbol{p}| - \omega \right)^2 f_1 \left( \tau, |\boldsymbol{k}| + \omega \right) + \left( |\boldsymbol{k}| + \omega \right)^2 f_1 \left( \tau, |\boldsymbol{p}| - \omega \right) - |\boldsymbol{p}|^2 f_1 \left( \tau, |\boldsymbol{k}| \right) - |\boldsymbol{k}|^2 f_1 \left( \tau, |\boldsymbol{p}| \right) \right] \\
                & + A_2^2 \left( \tau \right) \left[\left(|\boldsymbol{p}| - \omega\right)^2 \left(|\boldsymbol{k}| + \omega\right)^2 - |\boldsymbol{p}|^2 |\boldsymbol{k}|^2\right], \label{A4} \end{split} \end{align}

\noindent and the following shorthand notations are introduced:
\begin{align} \begin{split} & R_1 \left( \tau \right) \equiv \frac{2 \alpha^5 \left( \tau \right)}{\pi^2} \left[A_1^2 \left( \tau \right) - 2 A_0 \left( \tau \right) A_2 \left( \tau \right) + 5 A_1 \left( \tau \right) A_2 \left( \tau \right) \alpha \left( \tau \right) + 6 A_2^2 \left( \tau \right) \alpha^2 \left( \tau \right) \right], \\
                & R_2 \left( \tau \right) \equiv \frac{2 \alpha^4 \left( \tau \right)}{3 \pi^2} \left[-2 A_1^2 \left( \tau \right) + 4 A_0 \left( \tau \right) A_2 \left( \tau \right) - 5 A_1 \left( \tau \right) A_2 \left( \tau \right) \alpha \left( \tau \right) + 16 A_2^2 \left( \tau \right) \alpha^2 \left( \tau \right) \right], \\
                & R_3 \left( \tau \right) \equiv \frac{\alpha^3 \left( \tau \right)}{6 \pi^2} \left[A_1^2 \left( \tau \right) - 2 A_0 \left( \tau \right) A_2 \left( \tau \right) - 5 A_1 \left( \tau \right) A_2 \left( \tau \right) \alpha \left( \tau \right) - 44 A_2^2 \left( \tau \right) \alpha^2 \left( \tau \right) \right], \\
                & R_4 \left( \tau \right) \equiv \frac{A_2 \left( \tau \right) \alpha^3 \left( \tau \right)}{30 \pi^2} \left[5 A_1 \left( \tau \right) + 16 A_2 \left( \tau \right) \alpha \left( \tau \right) \right], \\
                & R_5 \left( \tau \right) \equiv \frac{A_2^2 \left( \tau \right) \alpha^2 \left( \tau \right)}{30 \pi^2}. \label{A5} \end{split} \end{align}
By substituting Eq.\eqref{A2} into the left-hand side of Eq.\eqref{220}, combining the result with Eq.\eqref{A3}, and matching the coefficients of identical powers of $p^0$ on both sides, we obtain a new set of ordinary differential equations that closely resembles Eq.\eqref{228}:
\begin{align} \begin{split} A_0' \left( \tau \right) = \frac{2 A_1^2 \left( \tau \right) \alpha^5 \left( \tau \right)}{\pi^2} - \frac{4 A_0 \left( \tau \right) A_2 \left( \tau \right) \alpha^5 \left( \tau \right)}{\pi^2} + \frac{10 A_1 \left( \tau \right) A_2 \left( \tau \right) \alpha^6 \left( \tau \right)}{\pi^2} + \frac{12 A_2^2 \left( \tau \right) \alpha^7 \left( \tau \right)}{\pi^2}, \label{A6} \end{split} \end{align}

\begin{align} \begin{split} 
                A_1' \left( \tau \right) + \frac{A_0 \left( \tau \right) \alpha' \left( \tau \right)}{\alpha^2 \left( \tau \right)} & = - \frac{4 A_1^2 \left( \tau \right) \alpha^4 \left( \tau \right)}{3 \pi^2} + \frac{8 A_0 \left( \tau \right) A_2 \left( \tau \right) \alpha^4 \left( \tau \right)}{3 \pi^2} - \frac{10 A_1 \left( \tau \right) A_2 \left( \tau \right) \alpha^5 \left( \tau \right)}{3 \pi^2} + \frac{32 A_2^2 \left( \tau \right) \alpha^6 \left( \tau \right)}{3 \pi^2}, \label{A7} \end{split} \end{align}

\begin{align} \begin{split} A_2' \left( \tau \right) + \frac{A_1 \left( \tau \right) \alpha' \left( \tau \right)}{\alpha^2 \left( \tau \right)} & = \frac{A_1^2 \left( \tau \right) \alpha^3 \left( \tau \right)}{6 \pi^2} - \frac{A_0 \left( \tau \right) A_2 \left( \tau \right) \alpha^3 \left( \tau \right)}{3 \pi^2}  - \frac{5 A_1 \left( \tau \right) A_2 \left( \tau \right) \alpha^4 \left( \tau \right)}{6 \pi^2} - \frac{22 A_2^2 \left( \tau \right) \alpha^5 \left( \tau \right)}{3 \pi^2}, \label{A8} \end{split} \end{align}

\begin{align} \begin{split} \frac{A_2 \left( \tau \right) \alpha' \left( \tau \right)}{\alpha^2 \left( \tau \right)} = \frac{A_1 \left( \tau \right) A_2 \left( \tau \right) \alpha^3 \left( \tau \right)}{6 \pi^2} +  \frac{8 A_2^2 \left( \tau \right) \alpha^4 \left( \tau \right)}{15 \pi^2}, \label{A9} \end{split} \end{align}

\begin{align} \begin{split} \frac{A_2^2 \left( \tau \right) \alpha^3 \left( \tau \right)}{30 \pi^2} = 0. \label{A10} \end{split} \end{align}
From Eq.\eqref{A10}, we directly obtain a nontrivial constraint
\begin{align} \begin{split} A_2 \left( \tau \right) = 0, \label{A11} \end{split} \end{align}
and find that Eq.\eqref{A2} reduces to Eq.\eqref{223}, Eqs.\eqref{A6}\textemdash \eqref{A8} reduce to Eq.\eqref{228}, and Eqs.\eqref{A9} and \eqref{A10} vanish identically. In other words,  $f_n \left( \tau, p^0 \right)$ for $n = 2$ is not a solution of the Boltzmann equation, unless it degenerates into the form corresponding to $n = 1$.

When $n = 3$, repeating the above steps yields another set of ordinary differential equations:
\begin{align} \begin{split} A_0' \left( \tau \right) = & \frac{2 A_1^2 \left( \tau \right) \alpha^5 \left( \tau \right)}{\pi^2} - \frac{4 A_0 \left( \tau \right) A_2 \left( \tau \right) \alpha^5 \left( \tau \right)}{\pi^2} + \frac{10 A_1 \left( \tau \right) A_2 \left( \tau \right) \alpha^6 \left( \tau \right)}{\pi^2} \\
                & + \frac{12 A_2^2 \left( \tau \right) \alpha^7 \left( \tau \right)}{\pi^2} - \frac{30 A_0 \left( \tau \right) A_3 \left( \tau \right) \alpha^6 \left( \tau \right)}{\pi^2} + \frac{36 A_1 \left( \tau \right) A_3 \left( \tau \right) \alpha^7 \left( \tau \right)}{\pi^2} \\
                & + \frac{84 A_2 \left( \tau \right) A_3 \left( \tau \right) \alpha^8 \left( \tau \right)}{\pi^2} + \frac{144 A_3^2 \left( \tau \right) \alpha^9 \left( \tau \right)}{\pi^2}, \label{A12} \end{split} \end{align}

\begin{align} \begin{split} A_1' \left( \tau \right) + \frac{A_0 \left( \tau \right) \alpha' \left( \tau \right)}{\alpha^2 \left( \tau \right)}  = & - \frac{4 A_1^2 \left( \tau \right) \alpha^4 \left( \tau \right)}{3 \pi^2} + \frac{8 A_0 \left( \tau \right) A_2 \left( \tau \right) \alpha^4 \left( \tau \right)}{3 \pi^2} - \frac{10 A_1 \left( \tau \right) A_2 \left( \tau \right) \alpha^5 \left( \tau \right)}{3 \pi^2} \\
                & + \frac{32 A_2^2 \left( \tau \right) \alpha^6 \left( \tau \right)}{3 \pi^2} + \frac{10 A_0 \left( \tau \right) A_3 \left( \tau \right) \alpha^5 \left( \tau \right)}{\pi^2} - \frac{28 A_1 \left( \tau \right) A_3 \left( \tau \right) \alpha^6 \left( \tau \right)}{\pi^2} \\
                & + \frac{76 A_2 \left( \tau \right) A_3 \left( \tau \right) \alpha^7 \left( \tau \right)}{\pi^2} + \frac{132 A_3^2 \left( \tau \right) \alpha^8 \left( \tau \right)}{\pi^2}, \label{A13} \end{split} \end{align}

\begin{align} \begin{split} A_2' \left( \tau \right) + \frac{A_1 \left( \tau \right) \alpha' \left( \tau \right)}{\alpha^2 \left( \tau \right)}  = & \frac{A_1^2 \left( \tau \right) \alpha^3 \left( \tau \right)}{6 \pi^2} - \frac{A_0 \left( \tau \right) A_2 \left( \tau \right) \alpha^3 \left( \tau \right)}{3 \pi^2} - \frac{5 A_1 \left( \tau \right) A_2 \left( \tau \right) \alpha^4 \left( \tau \right)}{6 \pi^2} \\
                & - \frac{22 A_2^2 \left( \tau \right) \alpha^5 \left( \tau \right)}{3 \pi^2} + \frac{5 A_0 \left( \tau \right) A_3 \left( \tau \right) \alpha^4 \left( \tau \right)}{2 \pi^2} + \frac{8 A_0 \left( \tau \right) A_3 \left( \tau \right) \alpha^5 \left( \tau \right)}{\pi^2} \\
                & - \frac{26 A_2 \left( \tau \right) A_3 \left( \tau \right) \alpha^6 \left( \tau \right)}{\pi^2} + \frac{60 A_3^2 \left( \tau \right) \alpha^7 \left( \tau \right)}{\pi^2}, \label{A14} \end{split} \end{align}

\begin{align} \begin{split} A_3' \left( \tau \right) + \frac{A_2 \left( \tau \right) \alpha' \left( \tau \right)}{\alpha^2 \left( \tau \right)} = & \frac{A_1 \left( \tau \right) A_2 \left( \tau \right) \alpha^3 \left( \tau \right)}{6 \pi^2} +  \frac{8 A_2^2 \left( \tau \right) \alpha^4 \left( \tau \right)}{15 \pi^2} - \frac{A_0 \left( \tau \right) A_3 \left( \tau \right) \alpha^3 \left( \tau \right)}{2 \pi^2} \\
                & - \frac{7 A_1 \left( \tau \right) A_3 \left( \tau \right) \alpha^4 \left( \tau \right)}{5 \pi^2} - \frac{26 A_2 \left( \tau \right) A_3 \left( \tau \right) \alpha^5 \left( \tau \right)}{5 \pi^2} - \frac{42 A_3^2 \left( \tau \right) \alpha^6 \left( \tau \right)}{\pi^2}, \label{A15} \end{split} \end{align}

\begin{align} \begin{split} \frac{A_3 \left( \tau \right) \alpha' \left( \tau \right)}{\alpha^2 \left( \tau \right)} = \frac{A_2^2 \left( \tau \right) \alpha^3 \left( \tau \right)}{30 \pi^2} + \frac{A_1 \left( \tau \right) A_3 \left( \tau \right) \alpha^3 \left( \tau \right)}{10 \pi^2} + \frac{19 A_2 \left( \tau \right) A_3 \left( \tau \right) \alpha^4 \left( \tau \right)}{30 \pi^2} + \frac{2 A_3^2 \left( \tau \right) \alpha^5 \left( \tau \right)}{\pi^2}, \label{A16} \end{split} \end{align}

\begin{align} \begin{split} \frac{A_2 \left( \tau \right) A_3 \left( \tau \right) \alpha^3 \left( \tau \right)}{30 \pi^2} + \frac{11 A_3^2 \left( \tau \right) \alpha^4 \left( \tau \right)}{70 \pi^2} = 0, \label{A17} \end{split} \end{align}

\begin{align} \begin{split} \frac{A_3^2 \left( \tau \right) \alpha^3 \left( \tau \right)}{140 \pi^2} = 0. \label{A18} \end{split} \end{align}
Once again from Eq.\eqref{A18}, we arrive at
\begin{align} \begin{split} A_3 \left( \tau \right) = 0. \label{A19} \end{split} \end{align}
Therefore, $f_n \left( \tau, p^0 \right)$ with $n = 3$ still  degenerates into the form of $n = 1$ similarly.

Finally, we note that for the case of $n \geq 2$,  the term $\left( p^0 - \omega \right)^n \omega^n$ coming from  $\left( p^0 - \omega \right)^n \left( k^0 + \omega \right)^n$ in Eq.\eqref{A1} gives  the highest power of $p^0$ after integration, whose coefficients have the form of $A_n^2 \left( \tau \right) \alpha^3 \left( \tau \right) / \left( c_n \pi^2 \right)$ where $c_n$ is a constant. However, on the left-hand side of Eq.\eqref{21}, this highest-power term does not appear, resulting in
\begin{align} \begin{split} \frac{A_n^2 \left( \tau \right) \alpha^3 \left( \tau \right)}{c_n \pi^2} = 0, \label{A20} \end{split} \end{align}
and
\begin{align} \begin{split} A_n \left( \tau \right) = 0. \label{A21} \end{split} \end{align}
Consequently, both $f_n \left( \tau, p^0 \right)$  and its associated system of ordinary differential equations reduce to the same form as previously analyzed, ultimately degenerating into Eq.\eqref{223} and Eq.\eqref{228}, respectively.

\section{DETAILS ON THE DETERMINATION OF \texorpdfstring{$A_0(\tau)$}{A0(tau)} AND \texorpdfstring{$A_1(\tau)$}{A1(tau)}}
\label{Determination}
{
In this appendix, we provide the detailed derivation for the coefficients $A_0\left(\tau\right)$ and $A_1\left(\tau\right)$ in terms of the conserved particle density $n_0$ and energy density $e_0$. Furthermore, we demonstrate the consistency of the resulting expressions with the system of ordinary differential equations \eqref{228} and show how the solution naturally recovers the classical Maxwell-Boltzmann distribution in the long-time limit.

Substituting the BKW solution \eqref{223} into the conservation condition Eq.\eqref{229} and evaluating the integrals in spherical coordinates, we obtain
\begin{align}
                n_0 & = \frac{1}{2 \pi^2} \int_0^\infty e^{- \frac{p}{\alpha \left( \tau \right)}} \left( A_0 \left( \tau \right) p^2 + A_1 \left( \tau \right) p^3 \right) \mathrm{d} p \nonumber \\
                & = \frac{A_0 \left( \tau \right)}{2 \pi^2} \int_0^\infty e^{- \frac{p}{\alpha \left( \tau \right)}} p^2 \mathrm{d} p + \frac{A_1 \left( \tau \right)}{2 \pi^2} \int_0^\infty e^{- \frac{p}{\alpha \left( \tau \right)}} p^3 \mathrm{d} p, \label{B1} \\
                e_0 & = \frac{1}{2 \pi^2} \int_0^\infty e^{- \frac{p}{\alpha \left( \tau \right)}} \left( A_0 \left( \tau \right) p^3 + A_1 \left( \tau \right) p^4 \right) \mathrm{d} p \nonumber \\
                & = \frac{A_0 \left( \tau \right)}{2 \pi^2} \int_0^\infty e^{- \frac{p}{\alpha \left( \tau \right)}} p^3 \mathrm{d} p + \frac{A_1 \left( \tau \right)}{2 \pi^2} \int_0^\infty e^{- \frac{p}{\alpha \left( \tau \right)}} p^4 \mathrm{d} p, \label{B2} 
\end{align}
where we still use the relationship $p^0 = p = |\boldsymbol{p}|$ for massless particles. Recalling the standard integral for integers $m \geq 0$,
\begin{align} \begin{split} \int_0^\infty e^{- \frac{p}{\alpha \left( \tau \right)}} p^m \mathrm{d} p = m! \  \alpha^{m + 1} \left( \tau \right), \label{B3} \end{split} \end{align}
Eq.\eqref{B1} and Eq.\eqref{B2} can then be evaluated as
\begin{align}
                n_0 & = \frac{1}{\pi^2} \left( A_0 \left( \tau \right) \alpha^3 \left( \tau \right) + 3 A_1 \left( \tau \right) \alpha^4 \left( \tau \right) \right), \label{B4} \\
                e_0 & = \frac{3}{\pi^2} \left( A_0 \left( \tau \right) \alpha^4 \left( \tau \right) + 4 A_1 \left( \tau \right) \alpha^5 \left( \tau \right) \right). \label{B5} 
\end{align}
Then we can express $A_0(\tau)$ and $A_1(\tau)$ in terms of $n_0$, $e_0$, and $\alpha(\tau)$,
\begin{align} \begin{split} A_0 \left( \tau \right) = \frac{\pi^2 \left( -e_0 + 4 n_0 \alpha \left( \tau \right) \right)}{\alpha^4 \left( \tau \right)}, \quad A_1 \left( \tau \right) = \frac{\pi^2 \left( e_0 - 3 n_0 \alpha \left( \tau \right) \right)}{3 \alpha^5 \left( \tau \right)}. \label{B6} \end{split} \end{align}
This reproduces Eq.\eqref{230} presented in the main text. Taking the derivatives of $A_0(\tau)$ and $A_1(\tau)$ with respect to $\tau$ yields
\begin{align} \begin{split} A_0^\prime \left( \tau \right) = \frac{4 \pi^2 \left( e_0 -3 n_0 \alpha \left( \tau \right) \right) \alpha^\prime \left( \tau \right)}{\alpha^5 \left( \tau \right)}, \quad A_1^\prime \left( \tau \right) = \frac{\pi^2 \left( -5 e_0 + 12 n_0 \alpha \left( \tau \right) \right) \alpha^\prime \left( \tau \right)}{3 \alpha^6 \left( \tau \right)}. \label{B7} \end{split} \end{align}
Rearranging Eq.\eqref{228} so that all terms are on the left-hand side, and substituting Eq.\eqref{B6} and Eq.\eqref{B7}, we find
\begin{align} \begin{split} 
                & A_0' \left( \tau \right) - \frac{2 A_1^2 \left( \tau \right) \alpha^5 \left( \tau \right)}{\pi^2} = - \frac{2 \pi^2}{9 \alpha^5 \left( \tau \right)} \left( e_0 - 3 n_0 \alpha \left( \tau \right) \right) \left( e_0 - 3 n_0 \alpha \left( \tau \right) -18 \alpha^\prime \left( \tau \right) \right), \\
                & A_1' \left( \tau \right) + \frac{A_0 \left( \tau \right) \alpha' \left( \tau \right)}{\alpha^2 \left( \tau \right)} + \frac{4 A_1^2 \left( \tau \right) \alpha^4 \left( \tau \right)}{3 \pi^2} = \frac{4 \pi^2}{27 \alpha^6 \left( \tau \right)} \left( e_0 - 3 n_0 \alpha \left( \tau \right) \right) \left( e_0 - 3 n_0 \alpha \left( \tau \right) -18 \alpha^\prime \left( \tau \right) \right), \\
                & \frac{A_1 \left( \tau \right) \alpha' \left( \tau \right)}{\alpha^2 \left( \tau \right)} - \frac{A_1^2 \left( \tau \right) \alpha^3 \left( \tau \right)}{6 \pi^2} = -\frac{\pi^2}{54 \alpha^7 \left( \tau \right)} \left( e_0 - 3 n_0 \alpha \left( \tau \right) \right) \left( e_0 - 3 n_0 \alpha \left( \tau \right) -18 \alpha^\prime \left( \tau \right) \right). \label{B8} \end{split} \end{align}
Upon substituting the explicit form of $\alpha(\tau)$ from Eq.\eqref{231}, all three expressions vanish identically. This confirms that the system in Eq.\eqref{228} is fully satisfied, as expected, since Eq.\eqref{231} was originally derived by substituting Eq.\eqref{230} into Eq.\eqref{228}.

Furthermore, Eq.\eqref{228} reveals a trivial steady-state solution for $\alpha(\tau)$, namely $\alpha(\tau) \equiv \alpha = \frac{e_0}{3 n_0}$. This coincides with the asymptotic limit of Eq.\eqref{231} as $\tau \to \infty$, which corresponds to the equilibrium state. In this limit, $A_0(\tau)$ and $A_1(\tau)$ become
\begin{equation}A_0 = \frac{27 \pi^2 n_0^4}{e_0^3}, \quad A_1 = 0.\end{equation}
Consequently, the distribution function reduces to
\begin{align} \begin{split} f_{\text{eq}} \left( p^0 \right) = \frac{27 \pi^2 n_0^4}{e_0^3} e^{- \frac{3 n_0 p^0}{e_0}}, \label{B10} \end{split} \end{align}
which is precisely the Maxwell-Boltzmann distribution.
}
\end{appendix}

\bibliographystyle{apsrev4-2}
\bibliography{main}

\begin{thebibliography}{31}%
\makeatletter
\providecommand \@ifxundefined [1]{%
 \@ifx{#1\undefined}
}%
\providecommand \@ifnum [1]{%
 \ifnum #1\expandafter \@firstoftwo
 \else \expandafter \@secondoftwo
 \fi
}%
\providecommand \@ifx [1]{%
 \ifx #1\expandafter \@firstoftwo
 \else \expandafter \@secondoftwo
 \fi
}%
\providecommand \natexlab [1]{#1}%
\providecommand \enquote  [1]{``#1''}%
\providecommand \bibnamefont  [1]{#1}%
\providecommand \bibfnamefont [1]{#1}%
\providecommand \citenamefont [1]{#1}%
\providecommand \href@noop [0]{\@secondoftwo}%
\providecommand \href [0]{\begingroup \@sanitize@url \@href}%
\providecommand \@href[1]{\@@startlink{#1}\@@href}%
\providecommand \@@href[1]{\endgroup#1\@@endlink}%
\providecommand \@sanitize@url [0]{\catcode `\\12\catcode `\$12\catcode `\&12\catcode `\#12\catcode `\^12\catcode `\_12\catcode `\%12\relax}%
\providecommand \@@startlink[1]{}%
\providecommand \@@endlink[0]{}%
\providecommand \url  [0]{\begingroup\@sanitize@url \@url }%
\providecommand \@url [1]{\endgroup\@href {#1}{\urlprefix }}%
\providecommand \urlprefix  [0]{URL }%
\providecommand \Eprint [0]{\href }%
\providecommand \doibase [0]{https://doi.org/}%
\providecommand \selectlanguage [0]{\@gobble}%
\providecommand \bibinfo  [0]{\@secondoftwo}%
\providecommand \bibfield  [0]{\@secondoftwo}%
\providecommand \translation [1]{[#1]}%
\providecommand \BibitemOpen [0]{}%
\providecommand \bibitemStop [0]{}%
\providecommand \bibitemNoStop [0]{.\EOS\space}%
\providecommand \EOS [0]{\spacefactor3000\relax}%
\providecommand \BibitemShut  [1]{\csname bibitem#1\endcsname}%
\let\auto@bib@innerbib\@empty
\bibitem [{\citenamefont {Heinz}(1985)}]{Heinz:1984yq}%
  \BibitemOpen
  \bibfield  {author} {\bibinfo {author} {\bibfnamefont {U.~W.}\ \bibnamefont {Heinz}},\ }\href {https://doi.org/10.1016/0003-4916(85)90336-7} {\bibfield  {journal} {\bibinfo  {journal} {Annals Phys.}\ }\textbf {\bibinfo {volume} {161}},\ \bibinfo {pages} {48} (\bibinfo {year} {1985})}\BibitemShut {NoStop}%
\bibitem [{\citenamefont {Heinz}(1986)}]{Heinz:1985qe}%
  \BibitemOpen
  \bibfield  {author} {\bibinfo {author} {\bibfnamefont {U.~W.}\ \bibnamefont {Heinz}},\ }\href {https://doi.org/10.1016/0003-4916(86)90114-4} {\bibfield  {journal} {\bibinfo  {journal} {Annals Phys.}\ }\textbf {\bibinfo {volume} {168}},\ \bibinfo {pages} {148} (\bibinfo {year} {1986})}\BibitemShut {NoStop}%
\bibitem [{\citenamefont {Bass}\ \emph {et~al.}(1998)\citenamefont {Bass} \emph {et~al.}}]{Bass:1998ca}%
  \BibitemOpen
  \bibfield  {author} {\bibinfo {author} {\bibfnamefont {S.~A.}\ \bibnamefont {Bass}} \emph {et~al.},\ }\href {https://doi.org/10.1016/S0146-6410(98)00058-1} {\bibfield  {journal} {\bibinfo  {journal} {Prog. Part. Nucl. Phys.}\ }\textbf {\bibinfo {volume} {41}},\ \bibinfo {pages} {255} (\bibinfo {year} {1998})},\ \Eprint {https://arxiv.org/abs/nucl-th/9803035} {arXiv:nucl-th/9803035} \BibitemShut {NoStop}%
\bibitem [{\citenamefont {Arnold}\ \emph {et~al.}(2000)\citenamefont {Arnold}, \citenamefont {Moore},\ and\ \citenamefont {Yaffe}}]{Arnold:2000dr}%
  \BibitemOpen
  \bibfield  {author} {\bibinfo {author} {\bibfnamefont {P.~B.}\ \bibnamefont {Arnold}}, \bibinfo {author} {\bibfnamefont {G.~D.}\ \bibnamefont {Moore}},\ and\ \bibinfo {author} {\bibfnamefont {L.~G.}\ \bibnamefont {Yaffe}},\ }\href {https://doi.org/10.1088/1126-6708/2000/11/001} {\bibfield  {journal} {\bibinfo  {journal} {JHEP}\ }\textbf {\bibinfo {volume} {2000}}\bibfield  {number} {\bibinfo  {number} { (11)},\ \bibinfo {pages} {001}},\ }\Eprint {https://arxiv.org/abs/hep-ph/0010177} {arXiv:hep-ph/0010177} \BibitemShut {NoStop}%
\bibitem [{\citenamefont {Molnar}\ and\ \citenamefont {Gyulassy}(2002)}]{Molnar:2001ux}%
  \BibitemOpen
  \bibfield  {author} {\bibinfo {author} {\bibfnamefont {D.}~\bibnamefont {Molnar}}\ and\ \bibinfo {author} {\bibfnamefont {M.}~\bibnamefont {Gyulassy}},\ }\href {https://doi.org/10.1016/S0375-9474(01)01224-6} {\bibfield  {journal} {\bibinfo  {journal} {Nucl. Phys. A}\ }\textbf {\bibinfo {volume} {697}},\ \bibinfo {pages} {495} (\bibinfo {year} {2002})},\ \bibinfo {note} {[Erratum: Nucl.Phys.A 703, 893--894 (2002)]},\ \Eprint {https://arxiv.org/abs/nucl-th/0104073} {arXiv:nucl-th/0104073} \BibitemShut {NoStop}%
\bibitem [{\citenamefont {Xu}\ and\ \citenamefont {Greiner}(2005)}]{Xu:2004mz}%
  \BibitemOpen
  \bibfield  {author} {\bibinfo {author} {\bibfnamefont {Z.}~\bibnamefont {Xu}}\ and\ \bibinfo {author} {\bibfnamefont {C.}~\bibnamefont {Greiner}},\ }\href {https://doi.org/10.1103/PhysRevC.71.064901} {\bibfield  {journal} {\bibinfo  {journal} {Phys. Rev. C}\ }\textbf {\bibinfo {volume} {71}},\ \bibinfo {pages} {064901} (\bibinfo {year} {2005})},\ \Eprint {https://arxiv.org/abs/hep-ph/0406278} {arXiv:hep-ph/0406278} \BibitemShut {NoStop}%
\bibitem [{\citenamefont {Denicol}\ \emph {et~al.}(2012)\citenamefont {Denicol}, \citenamefont {Niemi}, \citenamefont {Molnar},\ and\ \citenamefont {Rischke}}]{Denicol:2012cn}%
  \BibitemOpen
  \bibfield  {author} {\bibinfo {author} {\bibfnamefont {G.~S.}\ \bibnamefont {Denicol}}, \bibinfo {author} {\bibfnamefont {H.}~\bibnamefont {Niemi}}, \bibinfo {author} {\bibfnamefont {E.}~\bibnamefont {Molnar}},\ and\ \bibinfo {author} {\bibfnamefont {D.~H.}\ \bibnamefont {Rischke}},\ }\href {https://doi.org/10.1103/PhysRevD.85.114047} {\bibfield  {journal} {\bibinfo  {journal} {Phys. Rev. D}\ }\textbf {\bibinfo {volume} {85}},\ \bibinfo {pages} {114047} (\bibinfo {year} {2012})},\ \bibinfo {note} {[Erratum: Phys.Rev.D 91, 039902 (2015)]},\ \Eprint {https://arxiv.org/abs/1202.4551} {arXiv:1202.4551 [nucl-th]} \BibitemShut {NoStop}%
\bibitem [{\citenamefont {Dodelson}(2003)}]{Dodelson:2003ft}%
  \BibitemOpen
  \bibfield  {author} {\bibinfo {author} {\bibfnamefont {S.}~\bibnamefont {Dodelson}},\ }\href@noop {} {\emph {\bibinfo {title} {{Modern Cosmology}}}}\ (\bibinfo  {publisher} {Academic Press},\ \bibinfo {address} {Amsterdam},\ \bibinfo {year} {2003})\BibitemShut {NoStop}%
\bibitem [{\citenamefont {Weinberg}(2008)}]{Weinberg:2008zzc}%
  \BibitemOpen
  \bibfield  {author} {\bibinfo {author} {\bibfnamefont {S.}~\bibnamefont {Weinberg}},\ }\href@noop {} {\emph {\bibinfo {title} {{Cosmology}}}}\ (\bibinfo  {publisher} {OUP Oxford},\ \bibinfo {year} {2008})\BibitemShut {NoStop}%
\bibitem [{\citenamefont {Boltzmann}(1872)}]{boltzmann1872boltzmann}%
  \BibitemOpen
  \bibfield  {author} {\bibinfo {author} {\bibfnamefont {L.}~\bibnamefont {Boltzmann}},\ }\href@noop {} {\bibfield  {journal} {\bibinfo  {journal} {Sitzungsberichte der Kaiserlichen Akademie der Wissenschaften, Mathematisch-Naturwissenschaftliche Classe}\ }\textbf {\bibinfo {volume} {66}},\ \bibinfo {pages} {275} (\bibinfo {year} {1872})},\ \bibinfo {note} {english translation available: ``Further Studies on the Thermal Equilibrium of Gas Molecules''}\BibitemShut {NoStop}%
\bibitem [{\citenamefont {Bobylev}(1976)}]{Bobylev1}%
  \BibitemOpen
  \bibfield  {author} {\bibinfo {author} {\bibfnamefont {A.~V.}\ \bibnamefont {Bobylev}},\ }\href@noop {} {\bibfield  {journal} {\bibinfo  {journal} {Sov. Phys. Dokl}\ }\textbf {\bibinfo {volume} {20}},\ \bibinfo {pages} {820} (\bibinfo {year} {1976})}\BibitemShut {NoStop}%
\bibitem [{\citenamefont {Krook}\ and\ \citenamefont {Wu}(1976)}]{KW2}%
  \BibitemOpen
  \bibfield  {author} {\bibinfo {author} {\bibfnamefont {M.}~\bibnamefont {Krook}}\ and\ \bibinfo {author} {\bibfnamefont {T.~T.}\ \bibnamefont {Wu}},\ }\href {https://doi.org/10.1103/PhysRevLett.36.1107} {\bibfield  {journal} {\bibinfo  {journal} {Phys. Rev. Lett.}\ }\textbf {\bibinfo {volume} {36}},\ \bibinfo {pages} {1107} (\bibinfo {year} {1976})}\BibitemShut {NoStop}%
\bibitem [{\citenamefont {Krook}\ and\ \citenamefont {Wu}(1977)}]{KW}%
  \BibitemOpen
  \bibfield  {author} {\bibinfo {author} {\bibfnamefont {M.}~\bibnamefont {Krook}}\ and\ \bibinfo {author} {\bibfnamefont {T.~T.}\ \bibnamefont {Wu}},\ }\href {https://doi.org/10.1063/1.861780} {\bibfield  {journal} {\bibinfo  {journal} {Phys. Fluids}\ }\textbf {\bibinfo {volume} {20}},\ \bibinfo {pages} {1589} (\bibinfo {year} {1977})}\BibitemShut {NoStop}%
\bibitem [{\citenamefont {Krupp}(1967)}]{dissertation}%
  \BibitemOpen
  \bibfield  {author} {\bibinfo {author} {\bibfnamefont {R.}~\bibnamefont {Krupp}},\ }\href@noop {} {\bibfield  {journal} {\bibinfo  {journal} {M.Sc. thesis, MIT}\ } (\bibinfo {year} {1967})}\BibitemShut {NoStop}%
\bibitem [{\citenamefont {Ernst}(1984)}]{exactreview}%
  \BibitemOpen
  \bibfield  {author} {\bibinfo {author} {\bibfnamefont {M.~H.}\ \bibnamefont {Ernst}},\ }\href@noop {} {\bibfield  {journal} {\bibinfo  {journal} {Journal of Statistical Physics}\ }\textbf {\bibinfo {volume} {34}},\ \bibinfo {pages} {1001} (\bibinfo {year} {1984})}\BibitemShut {NoStop}%
\bibitem [{\citenamefont {Bird}(1994)}]{bird1994molecular}%
  \BibitemOpen
  \bibfield  {author} {\bibinfo {author} {\bibfnamefont {G.~A.}\ \bibnamefont {Bird}},\ }\href@noop {} {\emph {\bibinfo {title} {Molecular Gas Dynamics and the Direct Simulation of Gas Flows}}},\ \bibinfo {edition} {2nd}\ ed.\ (\bibinfo  {publisher} {Clarendon Press},\ \bibinfo {address} {Oxford},\ \bibinfo {year} {1994})\BibitemShut {NoStop}%
\bibitem [{\citenamefont {Bazow}\ \emph {et~al.}(2016{\natexlab{a}})\citenamefont {Bazow}, \citenamefont {Denicol}, \citenamefont {Heinz}, \citenamefont {Martinez},\ and\ \citenamefont {Noronha}}]{Bazow:2015dha}%
  \BibitemOpen
  \bibfield  {author} {\bibinfo {author} {\bibfnamefont {D.}~\bibnamefont {Bazow}}, \bibinfo {author} {\bibfnamefont {G.~S.}\ \bibnamefont {Denicol}}, \bibinfo {author} {\bibfnamefont {U.}~\bibnamefont {Heinz}}, \bibinfo {author} {\bibfnamefont {M.}~\bibnamefont {Martinez}},\ and\ \bibinfo {author} {\bibfnamefont {J.}~\bibnamefont {Noronha}},\ }\href {https://doi.org/10.1103/PhysRevLett.116.022301} {\bibfield  {journal} {\bibinfo  {journal} {Phys. Rev. Lett.}\ }\textbf {\bibinfo {volume} {116}},\ \bibinfo {pages} {022301} (\bibinfo {year} {2016}{\natexlab{a}})},\ \Eprint {https://arxiv.org/abs/1507.07834} {arXiv:1507.07834 [hep-ph]} \BibitemShut {NoStop}%
\bibitem [{\citenamefont {Bazow}\ \emph {et~al.}(2016{\natexlab{b}})\citenamefont {Bazow}, \citenamefont {Denicol}, \citenamefont {Heinz}, \citenamefont {Martinez},\ and\ \citenamefont {Noronha}}]{Bazow:2016oky}%
  \BibitemOpen
  \bibfield  {author} {\bibinfo {author} {\bibfnamefont {D.}~\bibnamefont {Bazow}}, \bibinfo {author} {\bibfnamefont {G.~S.}\ \bibnamefont {Denicol}}, \bibinfo {author} {\bibfnamefont {U.}~\bibnamefont {Heinz}}, \bibinfo {author} {\bibfnamefont {M.}~\bibnamefont {Martinez}},\ and\ \bibinfo {author} {\bibfnamefont {J.}~\bibnamefont {Noronha}},\ }\href {https://doi.org/10.1103/PhysRevD.94.125006} {\bibfield  {journal} {\bibinfo  {journal} {Phys. Rev. D}\ }\textbf {\bibinfo {volume} {94}},\ \bibinfo {pages} {125006} (\bibinfo {year} {2016}{\natexlab{b}})},\ \Eprint {https://arxiv.org/abs/1607.05245} {arXiv:1607.05245 [hep-ph]} \BibitemShut {NoStop}%
\bibitem [{\citenamefont {Hu}(2025)}]{Hu:2024utr}%
  \BibitemOpen
  \bibfield  {author} {\bibinfo {author} {\bibfnamefont {J.}~\bibnamefont {Hu}},\ }\href {https://doi.org/10.1007/JHEP07(2025)066} {\bibfield  {journal} {\bibinfo  {journal} {JHEP}\ }\textbf {\bibinfo {volume} {2025}}\bibfield  {number} {\bibinfo  {number} { (07)},\ \bibinfo {pages} {066}},\ }\Eprint {https://arxiv.org/abs/2411.16448} {arXiv:2411.16448 [hep-ph]} \BibitemShut {NoStop}%
\bibitem [{\citenamefont {Groot}\ \emph {et~al.}(1980)\citenamefont {Groot}, \citenamefont {Leeuwen}, \citenamefont {van Weert},\ and\ \citenamefont {Weert}}]{DeGroot:1980dk}%
  \BibitemOpen
  \bibfield  {author} {\bibinfo {author} {\bibfnamefont {S.}~\bibnamefont {Groot}}, \bibinfo {author} {\bibfnamefont {W.}~\bibnamefont {Leeuwen}}, \bibinfo {author} {\bibfnamefont {C.}~\bibnamefont {van Weert}},\ and\ \bibinfo {author} {\bibfnamefont {C.}~\bibnamefont {Weert}},\ }\href {https://books.google.co.jp/books?id=wkZ-AAAAIAAJ} {\emph {\bibinfo {title} {Relativistic Kinetic Theory: Principles and Applications}}}\ (\bibinfo  {publisher} {North-Holland Publishing Company},\ \bibinfo {year} {1980})\BibitemShut {NoStop}%
\bibitem [{\citenamefont {Hu}\ and\ \citenamefont {Shi}(2022)}]{Hu:2022vph}%
  \BibitemOpen
  \bibfield  {author} {\bibinfo {author} {\bibfnamefont {J.}~\bibnamefont {Hu}}\ and\ \bibinfo {author} {\bibfnamefont {S.}~\bibnamefont {Shi}},\ }\href {https://doi.org/10.1103/PhysRevD.106.014007} {\bibfield  {journal} {\bibinfo  {journal} {Phys. Rev. D}\ }\textbf {\bibinfo {volume} {106}},\ \bibinfo {pages} {014007} (\bibinfo {year} {2022})},\ \Eprint {https://arxiv.org/abs/2204.10100} {arXiv:2204.10100 [hep-ph]} \BibitemShut {NoStop}%
\bibitem [{\citenamefont {Denicol}\ and\ \citenamefont {Rischke}(2021)}]{Denicol:2021icf}%
  \BibitemOpen
  \bibfield  {author} {\bibinfo {author} {\bibfnamefont {G.~S.}\ \bibnamefont {Denicol}}\ and\ \bibinfo {author} {\bibfnamefont {D.~H.}\ \bibnamefont {Rischke}},\ }\href {https://doi.org/10.1007/978-3-030-82077-0} {\emph {\bibinfo {title} {Microscopic Foundations of Relativistic Fluid Dynamics}}},\ \bibinfo {series} {Lecture Notes in Physics}, Vol.\ \bibinfo {volume} {990}\ (\bibinfo  {publisher} {Springer},\ \bibinfo {address} {Cham},\ \bibinfo {year} {2021})\BibitemShut {NoStop}%
\bibitem [{\citenamefont {Berges}\ \emph {et~al.}(2008)\citenamefont {Berges}, \citenamefont {Rothkopf},\ and\ \citenamefont {Schmidt}}]{Berges:2008wm}%
  \BibitemOpen
  \bibfield  {author} {\bibinfo {author} {\bibfnamefont {J.}~\bibnamefont {Berges}}, \bibinfo {author} {\bibfnamefont {A.}~\bibnamefont {Rothkopf}},\ and\ \bibinfo {author} {\bibfnamefont {J.}~\bibnamefont {Schmidt}},\ }\href {https://doi.org/10.1103/PhysRevLett.101.041603} {\bibfield  {journal} {\bibinfo  {journal} {Phys. Rev. Lett.}\ }\textbf {\bibinfo {volume} {101}},\ \bibinfo {pages} {041603} (\bibinfo {year} {2008})},\ \Eprint {https://arxiv.org/abs/0803.0131} {arXiv:0803.0131 [hep-ph]} \BibitemShut {NoStop}%
\bibitem [{\citenamefont {Berges}\ \emph {et~al.}(2009)\citenamefont {Berges}, \citenamefont {Scheffler},\ and\ \citenamefont {Sexty}}]{Berges:2008mr}%
  \BibitemOpen
  \bibfield  {author} {\bibinfo {author} {\bibfnamefont {J.}~\bibnamefont {Berges}}, \bibinfo {author} {\bibfnamefont {S.}~\bibnamefont {Scheffler}},\ and\ \bibinfo {author} {\bibfnamefont {D.}~\bibnamefont {Sexty}},\ }\href {https://doi.org/10.1016/j.physletb.2009.10.032} {\bibfield  {journal} {\bibinfo  {journal} {Phys. Lett. B}\ }\textbf {\bibinfo {volume} {681}},\ \bibinfo {pages} {362} (\bibinfo {year} {2009})},\ \Eprint {https://arxiv.org/abs/0811.4293} {arXiv:0811.4293 [hep-ph]} \BibitemShut {NoStop}%
\bibitem [{\citenamefont {Pi{\~n}eiro~Orioli}\ \emph {et~al.}(2015)\citenamefont {Pi{\~n}eiro~Orioli}, \citenamefont {Boguslavski},\ and\ \citenamefont {Berges}}]{PineiroOrioli:2015cpb}%
  \BibitemOpen
  \bibfield  {author} {\bibinfo {author} {\bibfnamefont {A.}~\bibnamefont {Pi{\~n}eiro~Orioli}}, \bibinfo {author} {\bibfnamefont {K.}~\bibnamefont {Boguslavski}},\ and\ \bibinfo {author} {\bibfnamefont {J.}~\bibnamefont {Berges}},\ }\href {https://doi.org/10.1103/PhysRevD.92.025041} {\bibfield  {journal} {\bibinfo  {journal} {Phys. Rev. D}\ }\textbf {\bibinfo {volume} {92}},\ \bibinfo {pages} {025041} (\bibinfo {year} {2015})},\ \Eprint {https://arxiv.org/abs/1503.02498} {arXiv:1503.02498 [hep-ph]} \BibitemShut {NoStop}%
\bibitem [{\citenamefont {Mikheev}\ \emph {et~al.}(2023)\citenamefont {Mikheev}, \citenamefont {Siovitz},\ and\ \citenamefont {Gasenzer}}]{Mikheev:2023juq}%
  \BibitemOpen
  \bibfield  {author} {\bibinfo {author} {\bibfnamefont {A.~N.}\ \bibnamefont {Mikheev}}, \bibinfo {author} {\bibfnamefont {I.}~\bibnamefont {Siovitz}},\ and\ \bibinfo {author} {\bibfnamefont {T.}~\bibnamefont {Gasenzer}},\ }\href {https://doi.org/10.1140/epjs/s11734-023-00974-7} {\bibfield  {journal} {\bibinfo  {journal} {Eur. Phys. J. ST}\ }\textbf {\bibinfo {volume} {232}},\ \bibinfo {pages} {3393} (\bibinfo {year} {2023})},\ \Eprint {https://arxiv.org/abs/2304.12464} {arXiv:2304.12464 [cond-mat.quant-gas]} \BibitemShut {NoStop}%
\bibitem [{\citenamefont {Arnold}\ \emph {et~al.}(2003)\citenamefont {Arnold}, \citenamefont {Moore},\ and\ \citenamefont {Yaffe}}]{Arnold:2003}%
  \BibitemOpen
  \bibfield  {author} {\bibinfo {author} {\bibfnamefont {P.~B.}\ \bibnamefont {Arnold}}, \bibinfo {author} {\bibfnamefont {G.~D.}\ \bibnamefont {Moore}},\ and\ \bibinfo {author} {\bibfnamefont {L.~G.}\ \bibnamefont {Yaffe}},\ }\href {https://doi.org/10.1088/1126-6708/2003/05/051} {\bibfield  {journal} {\bibinfo  {journal} {JHEP}\ }\textbf {\bibinfo {volume} {2003}}\bibfield  {number} {\bibinfo  {number} { (05)},\ \bibinfo {pages} {051}},\ }\Eprint {https://arxiv.org/abs/hep-ph/0302165} {arXiv:hep-ph/0302165} \BibitemShut {NoStop}%
\bibitem [{\citenamefont {Soudi}(2021)}]{Soudi:2021}%
  \BibitemOpen
  \bibfield  {author} {\bibinfo {author} {\bibfnamefont {I.}~\bibnamefont {Soudi}},\ }\emph {\bibinfo {title} {{Energy loss and equilibration of a highly energetic parton in QCD plasmas}}},\ \href {https://doi.org/10.4119/unibi/2958574} {Ph.D. thesis},\ \bibinfo  {school} {Bielefeld U.} (\bibinfo {year} {2021})\BibitemShut {NoStop}%
\bibitem [{\citenamefont {Weinberg}(1972)}]{weinberg}%
  \BibitemOpen
  \bibfield  {author} {\bibinfo {author} {\bibfnamefont {S.}~\bibnamefont {Weinberg}},\ }\href@noop {} {\emph {\bibinfo {title} {Gravitation and Cosmology: Principles and Applications of the General Theory of Relativity}}}\ (\bibinfo  {publisher} {John Wiley \& Sons, New York},\ \bibinfo {year} {1972})\BibitemShut {NoStop}%
\bibitem [{\citenamefont {Bobylev}\ \emph {et~al.}(1996)\citenamefont {Bobylev}, \citenamefont {Caraffini},\ and\ \citenamefont {Spiga}}]{Bobylev:1996}%
  \BibitemOpen
  \bibfield  {author} {\bibinfo {author} {\bibfnamefont {A.~V.}\ \bibnamefont {Bobylev}}, \bibinfo {author} {\bibfnamefont {G.~L.}\ \bibnamefont {Caraffini}},\ and\ \bibinfo {author} {\bibfnamefont {G.}~\bibnamefont {Spiga}},\ }\href {https://doi.org/10.1063/1.531540} {\bibfield  {journal} {\bibinfo  {journal} {Journal of Mathematical Physics}\ }\textbf {\bibinfo {volume} {37}},\ \bibinfo {pages} {2787} (\bibinfo {year} {1996})}\BibitemShut {NoStop}%
\bibitem [{\citenamefont {Bobylev}(1993)}]{bobylev1993}%
  \BibitemOpen
  \bibfield  {author} {\bibinfo {author} {\bibfnamefont {A.~V.}\ \bibnamefont {Bobylev}},\ }\href@noop {} {\bibfield  {journal} {\bibinfo  {journal} {Mathematical Models and Methods in Applied Sciences}\ }\textbf {\bibinfo {volume} {3}},\ \bibinfo {pages} {443} (\bibinfo {year} {1993})}\BibitemShut {NoStop}%
\end{thebibliography}%

\end{document}